\documentclass[aps,prl,twocolumn,superscriptaddress,amsmath,amssymb,nofootinbib,floatfix]{revtex4-2}
\usepackage{graphicx}
\usepackage{bm}
\usepackage{xcolor}

\usepackage[colorlinks=true,allcolors=blue]{hyperref}
\usepackage{orcidlink}

\newcommand{\Hlep}{V_{\rm lep}}
\newcommand{\GF}{G_{\rm F}}

\begin{document}

\title{A muon matter potential tightens nucleosynthesis bounds on lepton flavor asymmetries}

\author{Van Que Tran\,\orcidlink{0000-0003-4643-4050}}
\email{vqtran@phys.ncts.ntu.edu.tw}
\affiliation{Physics Division, National Center for Theoretical Sciences, National Taiwan University, Taipei 106319, Taiwan}
\affiliation{Phenikaa Institute for Advanced Study, Phenikaa University, Nguyen Trac, Duong Noi, Hanoi 100000, Vietnam}

\date{\today}

\begin{abstract}
Charge neutrality fixes the sum of the electron and muon asymmetries in the early Universe, but not their difference.
When the muon- and electron-neutrino chemical potentials differ, the remaining muons give neutrinos a flavor-dependent matter potential that previous flavor evolution calculations neglected.
This potential locks the neutrino flavors until the temperature falls to about 5 MeV, when the thermal potential of electrons and positrons takes over and moves asymmetry into electron neutrinos.
For normal mass ordering, helium bounds on equal and opposite muon and tau lepton numbers therefore tighten by about an order of magnitude.
X-ray limits then restrict sterile-neutrino dark matter mixing with muon neutrinos to masses below about 18 keV, instead of about 50 keV.
\end{abstract}

\maketitle

How large can the lepton asymmetry of the Universe be, flavor by flavor?
Big-bang nucleosynthesis (BBN) answers through the electron-neutrino degeneracy $\xi_{\nu_e}=\mu_{\nu_e}/T$ at neutron freeze-out, which sets the helium abundance $Y_p$, and, together with the cosmic microwave background (CMB), through the energy density of neutrinos~\cite{Aver:2026dxv,Goldstein:2026iuu}.
At temperatures between about $T=20$ and $1\,$MeV, neutrino oscillations redistribute lepton number among the flavors~\cite{Savage:1990by,Dolgov:2002ab,Abazajian:2002qx,Wong:2002fa}.
They connect the measured quantities to the primordial asymmetries of the three flavors.
Recent solutions of the quantum kinetic equations with complete collision terms found directions in flavor space with zero total lepton number, along which little asymmetry reaches $\nu_e$~\cite{Domcke:2025lzg,Domcke:2025jiy}.
Along them, degeneracies of order one remain compatible with BBN.
These directions are now inputs for other physics.
Reference~\cite{Akita:2025txo} uses them to extend the parameter space of resonantly produced sterile-neutrino dark matter (DM), and Ref.~\cite{Gorbunov:2026ihq} to estimate its production for asymmetries above $0.1$.
Large flavored asymmetries also enter leptogenesis from lepton flavor violation~\cite{Mukaida:2021sgv} and a first-order QCD transition~\cite{Gao:2021nwz,Gonin:2026tyu}.

The flavor evolution calculations set the matter potential of charged lepton asymmetries to zero or to its baryonic value~\cite{Dolgov:2002ab,Abazajian:2002qx,Wong:2002fa,Pastor:2008ti,Mangano:2011ip,Johns:2016enc,Froustey:2021azz,Domcke:2025lzg}.
Reference~\cite{Savage:1990by} wrote this potential with the muon asymmetry and used charge neutrality to conclude that it is of the order of the baryon asymmetry.
Reference~\cite{Domcke:2025lzg} neglects it because below $T=20\,$MeV large asymmetries reside only in neutrinos.
Charge neutrality, however, constrains only the sum of the electron and muon asymmetries.
The share of lepton number held by charged leptons is known to matter for sterile-neutrino production near $T=100\,$MeV~\cite{Venumadhav:2015pla,Akita:2025txo}, and Ref.~\cite{Gorbunov:2026ihq} remarks that it could modify the allowed flavor directions.

In this Letter, we show that chemical equilibrium gives the muons that remain above temperatures of a few MeV an asymmetry whenever $\xi_{\nu_\mu}\neq\xi_{\nu_e}$, with the opposite asymmetry in electrons.
Through the charged-current interaction, these asymmetries give the neutrinos a flavor-dependent matter potential $\Hlep$.
The synchronization of large asymmetries does not suppress it.
We study the lepton asymmetry along the direction $L_\mu-L_\tau$, one of the least constrained, with $L_e=0$ and $L_\mu=-L_\tau$.\footnote{An asymmetry far above the baryon asymmetry needs physics beyond the Standard Model. Once created, it survives until oscillations act, because Standard Model interactions conserve each $L_\alpha$ and sphalerons conserve $L_\mu-L_\tau$.}
Here $L_\alpha=(\Delta n_{\nu_\alpha}+\Delta n_{\ell_\alpha})/s$ is the lepton number of flavor $\alpha$ per entropy density $s$, counting the neutrinos and the charged leptons $\ell_\alpha$, with $\Delta n\equiv n-\bar n$.
The potential $\Hlep$ locks the neutrino flavors until about $T=5\,$MeV. The thermal potential of $e^\pm$ then takes over and drives a level crossing that moves asymmetry into $\nu_e$.
Without $\Hlep$, the transfer to $\nu_e$ nearly cancels along this direction. 
For normal mass ordering, $\Hlep$ enhances $\xi_{\nu_e}$ at BBN about tenfold, tightening the bound on $L_\mu$ by about an order of magnitude. 
Consequently, sterile neutrinos mixing with $\nu_\mu$ require a mixing angle $\sin^22\theta$ roughly ten times larger, comparable to that required for flavor-universal asymmetries. 
X-ray constraints then lower the upper bound on sterile-neutrino DM mass from about $50\,$keV to about $18\,$keV.
The Supplemental Material gives the asymmetry variables, the potential and kinetic equations, the codes and their checks, the level crossing, the helium likelihood, the plane of zero total lepton number and sterile-neutrino dark matter.

\textit{The potential.}---At temperatures above a few MeV the reactions $\mu^-\leftrightarrow e^-\bar\nu_e\nu_\mu$, $\nu_\mu e^-\leftrightarrow\mu^-\nu_e$ and $e^-\bar\nu_e\leftrightarrow\mu^-\bar\nu_\mu$ keep the muons in chemical equilibrium. Muon decay alone exceeds the Hubble rate by more than $10^3$ down to $T=5\,$MeV, and the other reactions only speed up the equilibration. Equilibrium requires
\begin{equation}
\mu_\mu-\mu_e=\mu_{\nu_\mu}-\mu_{\nu_e},
\label{eq:cheq}
\end{equation}
where $\mu_{i}$ is the chemical potential of particle $i$.
Charge neutrality requires $\Delta n_e+\Delta n_\mu=n_p$, where $n_p$ is the proton density.
Electrons are abundant, so their chemical potential stays small.
The muon degeneracy is then close to $\xi_{\nu_\mu}-\xi_{\nu_e}$, and $\Delta n_\mu$ follows from the Fermi--Dirac distribution of a particle of mass $m_\mu$.
After a Fierz transformation, the charged-current interaction of $\nu_\alpha$ with the charged lepton $\ell_\alpha$ gives a matter potential $\sqrt2\GF\Delta n_{\ell_\alpha}$, where $\GF$ is the Fermi constant. In an isotropic, unpolarized medium only the vector-current density survives, so this holds for relativistic and non-relativistic leptons alike.
Neutral-current potentials are the same for all flavors, so the flavor-dependent part is
\begin{equation}
\Hlep=\sqrt2\,\GF\,{\rm diag}(n_p-\Delta n_\mu,\,\Delta n_\mu,\,0)
\label{eq:Hlep}
\end{equation}
in the flavor basis $(\nu_e,\nu_\mu,\nu_\tau)$.
It is odd under CP.
For a flavor-universal asymmetry, $\xi_{\nu_\mu}=\xi_{\nu_e}$, we have $\Delta n_\mu \simeq 0$, and $\Hlep$ reduces to the baryonic term.

The Boltzmann factor of the muons sets the size of $\Hlep$.
For $\xi_{\nu_\mu}-\xi_{\nu_e}=0.03$ at $T=10\,$MeV it exceeds the atmospheric oscillation frequency $\omega=\Delta m^2_{31}/2\langle E\rangle$, with $\langle E\rangle=3.15T$, by $3\times10^3$.
The charge-symmetric thermal potential of the muons~\cite{Notzold:1987ik} drives the muon-driven flavor transition near $T=12\,$MeV~\cite{Domcke:2025lzg}.
It carries the same Boltzmann factor, but $\Hlep$ exceeds it there by about $6\times10^5(\xi_{\nu_\mu}-\xi_{\nu_e})$.

\textit{Quantum kinetic equations.}---Each momentum mode $p$ is described by the $3\times3$ density matrices $\rho_p$ and $\bar\rho_p$ of neutrinos and antineutrinos in flavor space, whose diagonal entries are the occupation numbers.
In the convention of Ref.~\cite{Sigl:1993ctk} they obey
\begin{equation}
\begin{aligned}
i\dot\rho_p&=[\Omega_p+\Hlep+V_{\rm self},\,\rho_p]+iC_p,\\
i\dot{\bar\rho}_p&=[-\Omega_p+\Hlep+V_{\rm self},\,\bar\rho_p]+i\bar C_p,
\end{aligned}
\label{eq:qke}
\end{equation}
where $\Omega_p$ holds the vacuum and charge-symmetric thermal terms~\cite{Sigl:1993ctk,Notzold:1987ik}, and $C_p$ and $\bar C_p$ are the collision terms.
The self-interaction $V_{\rm self} = \sqrt2\GF D$ is set by the asymmetry matrix $D=\int_p(\rho_p-\bar\rho_p)$, with $\int_p\equiv\int d^3p/(2\pi)^3$, whose diagonal holds the asymmetries $\Delta n_{\nu_\alpha}$.

\textit{$\Hlep$ in the synchronized limit.}---The self-interaction of large asymmetries exceeds the spread of the vacuum oscillation frequencies over the thermal spectrum. 
All momentum modes then stay aligned and oscillate at one common frequency~\cite{Kostelecky:1994dt,Pastor:2001iu}. 
One might expect the self-interaction, which is much larger than $\Hlep$, to dominate the flavor evolution.
Subtracting the antineutrino equation in Eq.~\eqref{eq:qke} from the neutrino equation and integrating over momenta gives the equation for $D$.
In it the self-interaction drops out, because it is proportional to $D$ and commutes with it.

We illustrate this for two flavors $a$ and $b$ at one mean momentum.
We write $D=\tfrac12({\rm tr}\,D+\bm D\cdot\bm\sigma)$ and $S=\int_p(\rho_p+\bar\rho_p)=\tfrac12({\rm tr}\,S+\bm S\cdot\bm\sigma)$, 
where $\bm\sigma$ denotes the Pauli matrices and $\bm D$ and $\bm S$ are polarization vectors~\cite{Pastor:2001iu}, 
with the third component defining the flavor axis. 
Then $D_z=\Delta n_{\nu_a}-\Delta n_{\nu_b}$, and in the absence of collisions,
\begin{equation}
\dot{\bm D}=\omega\bm B\times\bm S+\lambda\bm L\times\bm D ,
\label{eq:Ddot}
\end{equation}
where $\bm L=(0,0,1)$, $\lambda=\sqrt2\GF(\Delta n_{\ell_a}-\Delta n_{\ell_b})$ arises from $\Hlep$, 
and $\bm B=(\sin2\theta_{ab},0,-\cos2\theta_{ab}+\Delta V_{\rm th}/\omega)$ 
incorporates the vacuum mixing angle $\theta_{ab}$ and the thermal potential difference $\Delta V_{\rm th}$. 
In the synchronized limit, $\bm S$ tracks $\bm D$, 
which precesses about $\omega(\bm S\cdot\bm D/D^2)\bm B+\lambda\bm L$. 
The vacuum and charge-symmetric thermal terms are therefore suppressed by the weight $\bm S\cdot\bm D/D^2$, whose size is about $0.4(\xi_{\nu_a}+\xi_{\nu_b})$ for small degeneracies, whereas the $\Hlep$ contribution remains unsuppressed. 
Flavor conversion is inhibited when $\lambda\gg\omega\,\bm S\cdot\bm D/D^2$. 
Since $\Hlep\propto\xi_{\nu_\mu}-\xi_{\nu_e}$, 
conversion begins at a temperature controlled by the muon Boltzmann suppression, nearly independently of the asymmetry magnitude, and by the $e^\pm$ part of $\Delta V_{\rm th}$. The three-flavor generalization is given in the Supplemental Material.

\textit{Numerical solutions.}---
From here on we specify the asymmetries by initial degeneracies. The initial degeneracy $\xi_{\nu_\alpha}$ is the degeneracy of $\nu_\alpha$ when it carries all of the conserved lepton number $L_\alpha$, so that $L_\alpha = \left(\xi_{\nu_\alpha}+\xi_{\nu_\alpha}^3/\pi^2\right)T^3/(6s)\,$.
With $\Hlep$ it is the physical $\nu_\alpha$ degeneracy at $T\simeq10\,$MeV, when the muons have decayed and the flavors are still locked. We read $\xi_{\nu_e}$ at $T=1.5\,$MeV and denote it by $\xi_{\nu_e}^{\rm BBN}$, its value at BBN.

We implemented $\Hlep$ in the momentum-averaged code 
COFLASY-C~\cite{Domcke:2025lzg,Domcke:2025jiy}. 
We also developed 
a momentum-resolved code that tracks the eigenbasis of the 
asymmetry matrix exactly in the synchronized limit. 
With full two-body collision integrals, the latter reproduces the neutrino decoupling computed with FortEPiaNO~\cite{Bennett:2020zkv} and, 
without $\Hlep$, agrees with the full neutrino transport 
of Ref.~\cite{Froustey:2024mgf} within $5\%$ in $\xi_{\nu_e}^{\rm BBN}$ 
for $|\xi_{\nu_\mu}|\le0.8$.
Including $\Hlep$ with normal mass ordering, the two codes agree 
on $\xi_{\nu_e}^{\rm BBN}$ within $10\%$ for $|\xi_{\nu_\mu}|\ge0.1$ 
of either sign of $\xi_{\nu_\mu}$ 
and near the BBN bound for $\xi_{\nu_\mu}<0$.
We further tested the synchronized limit in two ways, 
the adiabatic approximation of Ref.~\cite{Froustey:2021azz} 
and independent matrix evolution of each momentum mode with 
reduced self-interaction through the crossing.
Both agree with the momentum-resolved code within $9\%$ at the bounds, the latter only at the bound for $\xi_{\nu_\mu}<0$ and at the largest self-interaction strength reached. At the bound for $\xi_{\nu_\mu}>0$ the latter has not converged.
The collision terms with $\Hlep$ are tested only by the agreement of our own codes, and an independent calculation with $\Hlep$ and full collision terms would test them.

We present our results for normal neutrino mass ordering, using the oscillation parameters and parametrization of Ref.~\cite{Domcke:2025lzg}. In this parametrization, the asymmetry amplitude $A$ and angle $\phi$ specify the initial degeneracies for zero total lepton number.

\begin{figure}[!tb]
\includegraphics[width=\columnwidth]{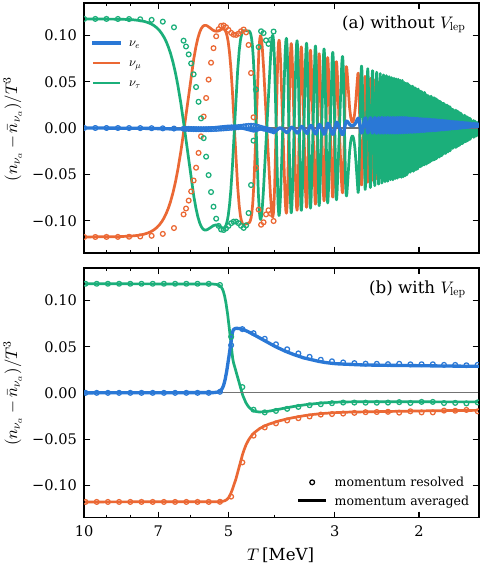}
\caption{Neutrino asymmetries $(n_{\nu_\alpha}-\bar n_{\nu_\alpha})/T^3$ as functions of temperature for normal mass ordering and amplitude $A=1$ along $L_\mu-L_\tau$, that is initial $\xi_{\nu_\alpha}=(0,-0.68,0.68)$, (a) without and (b) with $\Hlep$. Lines show the momentum-averaged solution and circles the momentum-resolved one with full collision integrals, both as instantaneous flavor asymmetries. In (a) the circles stop at $T=4\,$MeV.}
\label{fig:evol}
\end{figure}

\textit{Locking and level crossing.}---Figure~\ref{fig:evol} shows the evolution for $A=1$, that is $\xi_{\nu_\mu}=-\xi_{\nu_\tau}=-0.68$ and $\xi_{\nu_e}=0$.
At first both panels are frozen, for different reasons.
The vacuum term acts on the asymmetries only through the flavor differences of $\bm S$ in Eq.~\eqref{eq:Ddot}.
Here the densities of $\nu_\mu$ plus $\bar\nu_\mu$ and of $\nu_\tau$ plus $\bar\nu_\tau$ are equal, so the vacuum term cannot swap these two flavors directly and acts only through $\nu_e$.
Without $\Hlep$ the thermal potential of $e^\pm$ and collisional damping block this route.
When present, $\Hlep$ also holds the flavors. It is diagonal in flavor, like the asymmetry, so it exerts no torque and only keeps the asymmetry in place.
The two panels differ once these locks open.
Without $\Hlep$ this lock opens near $T=7\,$MeV, as the thermal potential falls toward the vacuum term and collisions slow down.
Nothing then makes the asymmetry precess about a definite axis in the $\mu\tau$ plane, so it cannot follow the changing eigenstates and is left in flavor states. This is the delayed, non-adiabatic response to the muon-driven transition of Ref.~\cite{Domcke:2025lzg}.
Because $\theta_{23}=47.6^\circ$, these are far from the vacuum eigenstates, and the $\nu_\mu$ and $\nu_\tau$ asymmetries oscillate with almost full amplitude, while $\nu_e$ hardly takes part.
Collisions act in the flavor basis and damp this oscillation, so its amplitude shrinks as the temperature falls.

With $\Hlep$ all flavors stay frozen down to about $T=5\,$MeV. The potential $\Hlep$ pushes the $\nu_e$ level above the $\nu_\tau$ level and the $\nu_\mu$ level below it, as the level diagram in the Supplemental Material shows.
In three flavors each pair of asymmetry eigenstates has its own weight, proportional to $\xi_{\nu_a}+\xi_{\nu_b}$ as in two flavors, positive here for the $\nu_e\nu_\tau$ pair. With this weight the thermal potential of $e^\pm$ and the vacuum term order the levels the other way.
Near $T=5\,$MeV $\Hlep$ falls below them, at about the temperature of the electron-driven transition of Ref.~\cite{Domcke:2025lzg}, where the $e^\pm$ potential falls below the vacuum term.
The $\nu_e$ level therefore falls through the $\nu_\tau$ level and, pushed by the $e^\pm$ potential, well below it. At this amplitude the asymmetry follows the instantaneous eigenstates and ends in vacuum eigenstates, which do not oscillate. The $\nu_\tau$ asymmetry ends in $\nu_1$ and the $\nu_\mu$ asymmetry in $\nu_3$, where $\nu_i$ are the mass eigenstates.
Without collisions $\nu_e$ would then hold $|U_{e1}|^2-|U_{e3}|^2\simeq0.66$ of the $\nu_\tau$ asymmetry, with $U$ the mixing matrix.
Collisions partly re-equilibrate the flavors, so about a quarter of the initial $\nu_\tau$ asymmetry ends in $\nu_e$ at BBN, against $2$ to $3\%$ without $\Hlep$. Without the $e^\pm$ potential the crossing would not be adiabatic, and the transfer would be about four times smaller.

\begin{figure}[!tb]
\includegraphics[width=\columnwidth]{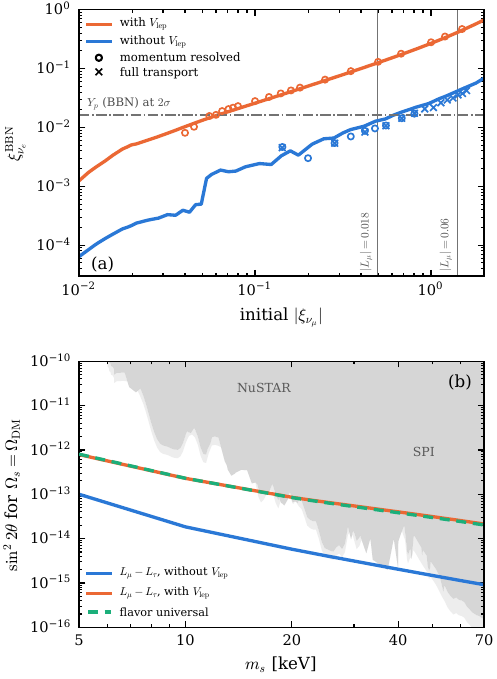}
\caption{(a) $\xi_{\nu_e}^{\rm BBN}$ along $L_\mu-L_\tau$, with initial $\xi_{\nu_e}=0$ and $\xi_{\nu_\mu}=-\xi_{\nu_\tau}<0$, for normal mass ordering, with $\Hlep$ (orange) and without (blue). Lines show the momentum-averaged solution. Circles show our momentum-resolved solution with full collision integrals. Crosses show the full neutrino transport of Ref.~\cite{Froustey:2024mgf} without $\Hlep$. The gray line marks where $Y_p$ alone leaves its $2\sigma$ range~\cite{Aver:2026dxv}, without the theory error on $\xi_{\nu_e}^{\rm BBN}$ and without $N_{\rm eff}$. Vertical lines mark $|L_\mu|=0.06$, the input of Ref.~\cite{Akita:2025txo}, and $|L_\mu|=0.018$, the reach quoted there for the Simons Observatory. (b) Mixing angle at which Majorana sterile neutrinos mixing with $\nu_\mu$ make up all DM, computed with the public code of Ref.~\cite{Akita:2025txo}. Lines are for $L_\mu-L_\tau$ at the BBN bound without (blue) and with (orange) $\Hlep$, and the dashed green line for flavor-universal asymmetries, $L_e=L_\mu=L_\tau$, at their BBN bound, all with our likelihood and the weaker sign of each bound. The region below each line is excluded by BBN. 
Shaded regions are excluded by NuSTAR at $95\%$ confidence~\cite{Krivonos:2024yvm,Neronov:2016wdd,Perez:2016tcq,Ng:2019gch,Roach:2019ctw,Roach:2022lgo} and by INTEGRAL/SPI at $3\sigma$~\cite{Fischer:2022pse}. Dark shading uses the weakest halo profile and the light bands span the profiles. 
}
\label{fig:main}
\end{figure}

\textit{Helium bound along $L_\mu-L_\tau$.}---Figure~\ref{fig:main}(a) shows $\xi_{\nu_e}^{\rm BBN}$ for $\xi_{\nu_\mu}<0$, and the Supplemental Material shows $\xi_{\nu_\mu}>0$.
With $\Hlep$, $\xi_{\nu_e}^{\rm BBN}$ is $0.2$ to $0.3$ times $|\xi_{\nu_\mu}|$ for $|\xi_{\nu_\mu}|\gtrsim0.04$, about ten times the value without it.
We combine the helium abundance of the Large Binocular Telescope $Y_p$ project, $Y_p=0.2458\pm0.0013$~\cite{Aver:2026dxv}, with the PRIMAT theory error~\cite{Pitrou:2018cgg}.
We add the effective number of neutrino species, $N_{\rm eff}=3.01\pm0.11$, from CMB and baryon acoustic oscillation data, which assume standard BBN helium~\cite{Goldstein:2026iuu}. 
We take into account a theory error on $\xi_{\nu_e}^{\rm BBN}$, which includes the difference between the momentum-averaged and momentum-resolved codes, an allowance for their collision approximations and a reproducibility error. 
We quote two-sided ranges at $95\%$ confidence for one parameter, with the baryon density fixed. With $\Hlep$, the allowed range shrinks from $-0.84<\xi_{\nu_\mu}<0.29$ to
 $-0.065<\xi_{\nu_\mu}<0.035$.
The two signs differ because the measured $Y_p$ lies below the standard prediction and favors $\xi_{\nu_e}^{\rm BBN}>0$. With data equal to the standard prediction the range shrinks from $-0.82<\xi_{\nu_\mu}<0.62$ to $-0.043<\xi_{\nu_\mu}<0.050$.
At the bounds the crossing is only partly adiabatic. Without collisions it would carry $64\%$ of the adiabatic transfer at the bound for $\xi_{\nu_\mu}<0$ and $32\%$ at the bound for $\xi_{\nu_\mu}>0$. Our codes follow this dynamics and do not assume adiabaticity. Computing the bounds with either code alone or with the full collision integrals moves them by up to about $11\%$.
In lepton number, our likelihood allows $-0.0023<L_\mu<0.0012$ with $\Hlep$, against $-0.032<L_\mu<0.010$ without.

Other statistical treatments, including the criteria of Ref.~\cite{Domcke:2025lzg}, and the full transport as the reference without $\Hlep$ give tightening factors between $6$ and $13$.
Below we use the bound for $\xi_{\nu_\mu}<0$ with our likelihood, the weaker one. 

\textit{Sterile-neutrino dark matter.}---Resonant production of keV sterile neutrinos needs a lepton asymmetry~\cite{Venumadhav:2015pla}. For zero total lepton number the asymmetry of the flavor they mix with sets the resonance.
The larger the asymmetry allowed by BBN, the smaller the mixing angle at which sterile neutrinos make up all the DM, and the heavier they can be before X-ray searches exclude that angle.
Reference~\cite{Akita:2025txo} showed that flavored asymmetries with zero total lepton number extend the allowed masses beyond those of flavor-universal asymmetries.
For mixing with $\nu_\mu$ it used $L_\mu=-L_\tau=0.06$, taken from Ref.~\cite{Domcke:2025lzg}, and noted that the Simons Observatory~\cite{SimonsObservatory:2018koc} could test $L_\mu\gtrsim0.018$.
Reference~\cite{Domcke:2025lzg} states that degeneracies of order one and larger are allowed along this direction.
Our bound without $\Hlep$ already excludes $L_\mu=0.06$, and with $\Hlep$ BBN also excludes $|L_\mu|=0.018$.

We compute the relic abundance with the public code of Ref.~\cite{Akita:2025txo}. 
We input our BBN bounds on $|L_\mu|$ to the code, which already includes the charged lepton asymmetries at the production temperatures, $T=100$ to $200\,$MeV. 
For comparison we computed with the same likelihood the BBN bound on a flavor-universal asymmetry, $L_e=L_\mu=L_\tau\equiv L_\alpha$. Such an asymmetry is not transferred between flavors, and $\Hlep$ reduces to the baryonic term. The bound, $L_\alpha<5.7\times10^{-4}$, is tighter than that on $|L_\mu|$ because $\nu_e$ carries its full asymmetry from the start.

Figure~\ref{fig:main}(b) shows the result.
Without $\Hlep$, the asymmetries allowed along $L_\mu-L_\tau$ give mixing angles $\sin^22\theta$ that are about one order of magnitude smaller than those of flavor-universal asymmetries.
With $\Hlep$ the angle rises by the same factor and comes within $5\%$ of the flavor-universal value at every mass we computed.
This closeness is partly a coincidence.
The matter potential that drives the resonance is proportional to $2L_\mu+L_e+L_\tau$~\cite{Akita:2025txo}, which is $L_\mu$ along $L_\mu-L_\tau$ and $4L_\alpha$ for a flavor-universal asymmetry. With $\Hlep$ about a quarter of $\xi_{\nu_\mu}$ reaches $\nu_e$, so the bound on $L_\mu$ is about four times the flavor-universal bound. The factor of four in the resonance potential is fixed, but the quarter is set by the collision rates.

We take as the mass limit the lightest mass at which the angle reaches an X-ray limit. With the median expected NuSTAR limit of Ref.~\cite{Krivonos:2024yvm} and the SPI limit, it falls from $m_s \simeq 52\,$keV without $\Hlep$ to about $18\,$keV with it, the same as for flavor-universal asymmetries.
The observed limits give $m_s\simeq16\,$keV with $\Hlep$ for every halo profile, and $25$ or $36\,$keV without it, where those limits fluctuate below their expectation~\cite{Krivonos:2024yvm} or come from a single straight segment of the drawn compilation.
Production uncertainties of tens of percent move the observed limit with $\Hlep$ to at most $m_s = 19\,$keV.
Above it, the angle with $\Hlep$ stays allowed only in windows narrower than $1\,$keV, up to $m_s\simeq21\,$keV.
For our spectra, the indicative Lyman-$\alpha$ 
bound~\cite{Akita:2025txo} lies at $m_s\simeq16.5\,$keV 
with $\Hlep$ and $18\,$keV without it, leaving little 
or no viable parameter space along the $L_\mu-L_\tau$ direction.
For these non-thermal spectra it is an estimate rather than a strict bound.

\textit{Other directions.}---Across the plane of zero total lepton number, $\Hlep$ changes the size of $\xi_{\nu_e}^{\rm BBN}$ only moderately for normal ordering. The minimal washout factor of a generic direction, the smallest ratio $A/|\xi_{\nu_e}^{\rm BBN}|$~\cite{Domcke:2025lzg}, falls from about $6$ to about $5$.
The sign of $\xi_{\nu_e}^{\rm BBN}$, however, reverses in $80$ to $90\%$ of all directions, which the momentum-resolved code confirms at five generic directions with $A=0.25$.
The potential $\Hlep$ therefore changes which flavor compositions raise the helium abundance and which lower it.
The tauphobic line $L_e=-L_\mu$ is where a first-order QCD transition needs the smallest asymmetries~\cite{Gao:2021nwz,Gao:2024fhm}.
For the sign relevant there, $\xi_{\nu_e}>0$, the observed bound hardly changes, because the low measured $Y_p$ already disfavors the sign of $\xi_{\nu_e}^{\rm BBN}$ found without $\Hlep$.
With $\Hlep$ the expected bound improves from $|\xi_{\nu_\mu}|<0.13$ to $0.037$. On this line $\xi_{\nu_e}=|\xi_{\nu_\mu}|$ initially, so a threshold three times the $\xi_{\nu_e}\simeq0.04$ of Ref.~\cite{Gao:2021nwz}, as Ref.~\cite{Gao:2024fhm} suggests, is excluded by both bounds.
Taking the weaker sign of each bound, the bound on $|L_\mu|$ along this line tightens from between $0.008$ and $0.018$ to between $0.002$ and $0.006$, depending on the statistics.
Narrow valleys, directions in which $\xi_{\nu_e}^{\rm BBN}$ passes through zero, allow large asymmetry amplitudes~\cite{Domcke:2025lzg}.
With $\Hlep$ they move by about $10^\circ$ in the angle $\phi$, and the valley along $L_\mu-L_\tau$ disappears.
For degeneracies of order one, the asymmetry amplitude $A\gtrsim1$, $\xi_{\nu_e}^{\rm BBN}$ near the shifted valleys changes erratically within a degree.
There two eigenvalues of the asymmetry matrix nearly meet, and our two codes disagree on the sign of $\xi_{\nu_e}^{\rm BBN}$ at $6$ of $21$ directions.
Whether allowed valleys survive at large asymmetry is therefore not settled.

\textit{Conclusions.}---Along $L_\mu-L_\tau$ the potential $\Hlep$ tightens the helium bounds by about an order of magnitude and limits sterile-neutrino DM mixing with $\nu_\mu$ to masses below about $18\,$keV. These results hold for normal ordering, and the valleys at large asymmetry are not resolved. 
For inverted ordering, $\Hlep$ also changes $\xi_{\nu_e}^{\rm BBN}$ about tenfold, but the final value depends on the phase of collective oscillations when collisions freeze out, and we quote no bound.
The potential $\Hlep$ follows from Standard Model physics alone.
It acts whenever the muon- and electron-neutrino degeneracies differ before BBN, and for normal ordering it is largest along the directions that were least constrained.
Results that rely on large flavored asymmetries surviving to BBN should be recomputed with it. Examples are the remaining directions used for sterile-neutrino production~\cite{Akita:2025txo,Gorbunov:2026ihq}, heavy neutral leptons in asymmetric backgrounds~\cite{Akita:2026gee} and the bound on the total lepton number~\cite{Domcke:2025jiy}, whose generic directions have $\xi_{\nu_\mu}\neq\xi_{\nu_e}$.

\begin{acknowledgments}
We thank the authors of Refs.~\cite{Domcke:2025lzg,Domcke:2025jiy,Akita:2025txo,Fischer:2022pse,Froustey:2024mgf,Bennett:2020zkv} for making their codes and data public. This work was supported in part by the National Science and Technology Council (NSTC), the
Ministry of Education (Higher Education Sprout Project NTU-114L104022-1) and the National Center for Theoretical Sciences of Taiwan.
Parts of the numerical work and of the drafting were assisted by an AI system (Claude, Anthropic). 
The author checked the results and is fully responsible for the work.
\end{acknowledgments}

\nocite{Planck:2018vyg}
\bibliography{refs}

\clearpage
\onecolumngrid
\setcounter{secnumdepth}{1}
\setcounter{section}{0}
\setcounter{equation}{0}
\setcounter{figure}{0}
\setcounter{table}{0}
\renewcommand{\thesection}{S\arabic{section}}
\renewcommand{\theequation}{S\arabic{equation}}
\renewcommand{\thefigure}{S\arabic{figure}}
\renewcommand{\thetable}{S\Roman{table}}
\renewcommand{\theHsection}{S\arabic{section}}
\renewcommand{\theHequation}{S\arabic{equation}}
\renewcommand{\theHfigure}{S\arabic{figure}}
\renewcommand{\theHtable}{S\Roman{table}}

\begin{center}
{\large\bfseries Supplemental Material}
\end{center}
This Supplemental Material gives the asymmetry variables, the equations and numerical methods behind the Letter, the checks of our codes, the level crossing, the helium likelihood, the plane of zero total lepton number and sterile-neutrino dark matter.
Equations, figures and tables of the Letter are cited as Eq.~(n) and Fig.~n, and those of this Supplemental Material carry the prefix S.

\section{Asymmetry variables and parametrization}\label{sec:param}
\textit{Variables.} The neutrino asymmetries are described by the degeneracies $\xi_{\nu_\alpha}=\mu_{\nu_\alpha}/T$, where $T$ is the comoving temperature, which falls as the inverse scale factor and equals the photon temperature before $e^\pm$ annihilation. For a Fermi--Dirac spectrum of one helicity, $\Delta n_{\nu_\alpha}=T^3(\xi_{\nu_\alpha}+\xi_{\nu_\alpha}^3/\pi^2)/6$. The charged leptons carry $\Delta n_{\ell_\alpha}$, fixed by chemical equilibrium and charge neutrality (Sec.~\ref{sec:eqs}). We read $\xi_{\nu_e}$ at $T=1.5\,$MeV and call it $\xi_{\nu_e}^{\rm BBN}$, its value at BBN. It is the oscillation-averaged asymmetry, the projection of $D$ onto the vacuum mass basis, which a dephasing of the modes below about $T=1.5\,$MeV leaves unchanged. Reading it at $T=0.8\,$MeV instead changes the bounds by at most about $10\%$ (Sec.~\ref{sec:bounds}). The baryon-to-photon ratio is fixed at $\eta_B=6.1\times10^{-10}$~\cite{Planck:2018vyg} throughout.

\textit{Lepton numbers.} Apart from oscillations, which start below $T\simeq20\,$MeV, Standard Model interactions conserve each flavor lepton number $L_\alpha=(\Delta n_{\nu_\alpha}+\Delta n_{\ell_\alpha})/s$, which counts the neutrinos and the charged leptons.
How $L_\alpha$ is shared between them changes with temperature. At $T=20\,$MeV the muons hold $11\%$ of $L_\mu$ for $\xi_{\nu_e}=0$, and below $T=10\,$MeV less than $0.2\%$. Taus are absent at these temperatures.
Once the muons are gone, below $T\simeq10\,$MeV, each $L_\alpha$ sits in the neutrinos, and for every flavor and every direction
\begin{equation}
\label{eq:Lalpha}
L_\alpha \simeq \left(\xi_{\nu_\alpha}+\frac{\xi_{\nu_\alpha}^3}{\pi^2}\right) \frac{T^3}{6s},
\end{equation}
with $6s/T^3=4\pi^2g_{*s}/15=28.3$ for $g_{*s}=10.75$, as in Eq.~(4.1) of Ref.~\cite{Domcke:2025lzg}. The relation assumes Fermi--Dirac spectra and neglects the entropy of the asymmetries themselves, a correction of order $\xi_{\nu_\alpha}^2$.

\textit{Initial degeneracies.} We specify each run by the initial degeneracies $\xi_{\nu_\alpha}$, defined by Eq.~\eqref{eq:Lalpha} as the degeneracies $\nu_\alpha$ would have if it carried all of $L_\alpha$, as in Refs.~\cite{Domcke:2025lzg,Gorbunov:2026ihq}.
With $\Hlep$ they equal the physical neutrino degeneracies at $T\simeq10\,$MeV, because by then the muons have decayed and $\Hlep$ keeps the flavors locked. In the four directions we checked the asymmetries change by less than $0.1\%$ between $T=20$ and $10\,$MeV.
Without $\Hlep$, as in Ref.~\cite{Domcke:2025lzg}, they are the degeneracies at $T=20\,$MeV.
The codes start at $T=20\,$MeV with the initial degeneracies. Starting them at any temperature between $T=30$ and $8\,$MeV changes $\xi_{\nu_e}^{\rm BBN}$ at the bound for $\xi_{\nu_\mu}<0$ by less than $0.03\%$. Physically, at $T=20\,$MeV the muons still hold part of $L_\mu$, and charge neutrality gives $\nu_e$ an equal asymmetry. Solving chemical equilibrium with the conserved $L_\mu$ and $L_e$, for $L_e=0$, gives a $\nu_\mu$ degeneracy $7\%$ below its initial value and $\xi_{\nu_e}\simeq0.11\,\xi_{\nu_\mu}$ there. Muon decay removes both differences by $T=10\,$MeV, without flavor exchange.
With $\Hlep$ this enters only through Eq.~\eqref{eq:dnmu} above $T\simeq10\,$MeV, where the initial $\xi_{\nu_\mu}-\xi_{\nu_e}$ exceeds the physical value by $21\%$ at $T=20\,$MeV and by $1.2\%$ at $T=12\,$MeV. There $\Hlep$ exceeds the vacuum term by more than $10^3$, and the scaling test of Sec.~\ref{sec:codes}, which changes $\Hlep$ far more and at all temperatures, bounds the effect.

\textit{Sterile-neutrino production.} Because each $L_\alpha$ is conserved between $T=200$ and $20\,$MeV, a bound on $L_\alpha$ from BBN applies at the production temperatures, up to the depletion by production itself (Sec.~\ref{sec:xray}). The code of Ref.~\cite{Akita:2025txo} solves $\Delta n_{\nu_\alpha}+\Delta n_{\ell_\alpha}=L_\alpha s$ with charge neutrality at each temperature, so it distributes $L_\alpha$ among neutrinos and charged leptons there. We pass it our bounds on $L_\alpha$.

We parametrize zero total lepton number as in Ref.~\cite{Domcke:2025lzg}, 
\begin{equation}
y_\alpha\equiv\xi_{\nu_\alpha}+\xi_{\nu_\alpha}^3/\pi^2
=A(\cos\phi\sin\vartheta,\sin\phi\sin\vartheta,\cos\vartheta), 
\end{equation}
for $\alpha \equiv (e,\mu,\tau)$
with $\tan\vartheta=-1/(\cos\phi+\sin\phi)$.
The amplitude is $A^2=\sum_\alpha y_\alpha^2$, and the angle $\vartheta$ enforces $\sum_\alpha y_\alpha=0$, that is zero total lepton number. The angle $\phi$ is the direction in this plane.
Adding $180^\circ$ to $\phi$ reverses all signs. Tauphobic asymmetries, $\xi_{\nu_e}=-\xi_{\nu_\mu}$ and $\xi_{\nu_\tau}=0$, are $\phi=135^\circ$ or $315^\circ$.

Most results of the Letter are for $L_\mu-L_\tau$ with $L_e=0$, that is $\xi_{\nu_e}=0$ and $\xi_{\nu_\mu}=-\xi_{\nu_\tau}$ for the initial degeneracies. Because Eq.~\eqref{eq:Lalpha} is the same for every flavor, $L_e=0$ and $L_\mu=-L_\tau$ then hold exactly for the conserved lepton numbers, up to the electrons that neutralize the protons. This direction is $\phi=90^\circ$ for $\xi_{\nu_\mu}>0$ and $\phi=270^\circ$ for $\xi_{\nu_\mu}<0$, with $A=\sqrt2\,|y_\mu|$. Section~\ref{sec:plane} covers the other directions.

\section{Potential and kinetic equations}\label{sec:eqs}

\textit{Muon asymmetry.} The muon decay rate, time dilated by $\langle m_\mu/E\rangle$, exceeds the Hubble rate by $6\times10^3$ at $T=10\,$MeV and by $2.5\times10^4$ at $T=5\,$MeV, so the muons stay in chemical equilibrium, shown in Eq.~(1).
In the neutrality condition $\Delta n_e+\Delta n_\mu=n_p$ we take $n_p=(11/4)\eta_Bn_\gamma X_p$, with the equilibrium proton fraction $X_p$.
The electron chemical potential is about $6\times10^{-3}T$ at $T=20\,$MeV for $\xi_{\nu_\mu}-\xi_{\nu_e}=0.1$, so the muon degeneracy is close to $\xi_{\nu_\mu}-\xi_{\nu_e}$, and the net muon density of two spin states is
\begin{equation}
\Delta n_\mu=\frac{2m_\mu^2T}{\pi^2}\sum_{k\ge1}\frac{(-1)^{k+1}}{k}\,K_2\!\left(\frac{km_\mu}{T}\right)\sinh\!\big[k(\xi_{\nu_\mu}-\xi_{\nu_e})\big].
\label{eq:dnmu}
\end{equation}
We solve Eqs.~(1) and \eqref{eq:dnmu} together with neutrality by iteration.
Table~\ref{tab:scales} compares $\sqrt2\GF\Delta n_\mu$ with the atmospheric vacuum frequency $\omega=\Delta m^2_{31}/2\langle E\rangle$, with $\langle E\rangle=3.15T$.
The baryonic term in Eq.~(2) becomes comparable to the vacuum term weighted as in Eq.~(4) of the Letter for small asymmetries. It changes $|\xi_{\nu_e}^{\rm BBN}|$ by $5\%$ at the bound for $\xi_{\nu_\mu}<0$, by $12\%$ at the bound for $\xi_{\nu_\mu}>0$ and by $1\%$ at $A=1$.

\begin{table}[t]
\caption{Ratio of $\sqrt2\GF\Delta n_\mu$ to $\omega$ for $\xi_{\nu_e}=0$ and several $\xi_{\nu_\mu}$.}
\label{tab:scales}
\begin{tabular}{lcccccc}
\hline\hline
$T$ [MeV] & 20 & 12 & 10 & 8 & 6 & 5\\
\hline
$\xi_{\nu_\mu}=0.003$ & $4\times10^5$ & $3\times10^3$ & $330$ & $13$ & $0.07$ & $0.001$\\
$\xi_{\nu_\mu}=0.03$ & $4\times10^6$ & $3\times10^4$ & $3\times10^3$ & $130$ & $0.7$ & $0.014$\\
$\xi_{\nu_\mu}=0.24$ & $3\times10^7$ & $2.5\times10^5$ & $3\times10^4$ & $10^3$ & $6$ & $0.11$\\
\hline\hline
\end{tabular}
\end{table}

\textit{Three flavors.} Write $D=N-\bar N$ for the asymmetry matrix, with $N=\int_p\rho_p$ and $\bar N=\int_p\bar\rho_p$ in the notation of the Letter, $\rho_p$ and $\bar\rho_p$ for the density matrices of mode $p$, and $\Omega_p$ for the vacuum plus charge-symmetric thermal Hamiltonian,
\begin{equation}
\Omega_p=\frac{UM^2U^\dagger}{2p}-\frac{2\sqrt2\GF p}{m_W^2}\,{\rm diag}\big(\varepsilon_e+P_e,\,\varepsilon_\mu+P_\mu,\,0\big).
\label{eq:Omega}
\end{equation}
Here $U$ is the PMNS mixing matrix, $M^2={\rm diag}(0,\Delta m^2_{21},\Delta m^2_{31})$, $m_W$ the $W$ mass, and $\varepsilon_\ell+P_\ell$ the energy density plus pressure of $\ell^\pm$. Following Ref.~\cite{Domcke:2025lzg}, we use $\Delta m^2_{21}=7.53\times10^{-5}\,$eV$^2$, $\theta_{12}=33.6^\circ$, $\theta_{13}=8.5^\circ$, $\delta_{\rm CP}=0$, and $\Delta m^2_{31}=2.534\times10^{-3}\,$eV$^2$, $\theta_{23}=47.6^\circ$ for normal ordering, and $\Delta m^2_{31}=-2.46\times10^{-3}\,$eV$^2$, $\theta_{23}=47.2^\circ$ for inverted ordering.
For relativistic $e^\pm$, the second term in~Eq.~\eqref{eq:Omega} is the thermal potential~\cite{Notzold:1987ik}.
In the convention of Eq.~(3), in which antineutrinos evolve with $-\Omega_p$ and $+(V_{\rm self}+\Hlep)$~\cite{Sigl:1993ctk}, one finds the exact identity
\begin{equation}
\dot D=-i\int_p[\Omega_p,\rho_p+\bar\rho_p]-i[\Hlep,D]+\int_p(C_p-\bar C_p),
\label{eq:Dmat}
\end{equation}
in which the self-interaction and the charge-symmetric neutrino-neutrino thermal term drop out. This is the three-flavor form of Eq.~(4).
When the self-interaction is large, $S=\int_p(\rho_p+\bar\rho_p)$ commutes with $D$ up to small corrections, so the vacuum term enters only through this combination, while $\Hlep$ acts on $D$ directly.

\section{Codes and checks}\label{sec:codes}

We solve Eq.~(3) of the Letter with two codes that treat the momentum dependence differently.
The momentum-averaged code follows one density matrix for all momenta. The momentum-resolved code follows every momentum mode in the synchronized limit.
Both codes add $\Hlep$ from Eqs.~(1), (2) and \eqref{eq:dnmu} at every step.

\textit{Momentum-averaged code.}
The code COFLASY~\cite{Domcke:2025lzg,Domcke:2025jiy} assumes that all momentum modes of the neutrinos share one Fermi--Dirac spectrum with zero chemical potential. This spectrum is multiplied by a momentum-independent matrix $r$ that carries the asymmetries and the flavor coherence, and likewise $\bar r$ for the antineutrinos.
Integrating Eq.~(3) over momenta then gives kinetic equations for these two $3\times3$ matrices, which the code solves together with the neutrino and photon temperatures.
The vacuum and thermal terms enter through averages over the common spectrum.
The collision terms are derived for this ansatz and include scattering on $e^\pm$, annihilation into $e^+e^-$ and neutrino self-interactions~\cite{Domcke:2025lzg}.
The ansatz does not allow the spectra of different modes to differ, so the code misses the momentum dependence of the oscillations. The momentum-resolved code below includes it.

We use the C++ version, COFLASY-C 0.1.058~\cite{Domcke:2025lzg,Domcke:2025jiy}.
At each step we read the neutrino degeneracies off the diagonal of its asymmetry matrix, solve Eqs.~(1) and \eqref{eq:dnmu} with charge neutrality, and add $-i[\Hlep,r]$ and $-i[\Hlep,\bar r]$ to the equations for $r$ and $\bar r$.
COFLASY-C writes antineutrinos in the conjugate convention, in which the vacuum and charge-symmetric terms change sign and the asymmetry terms do not, so $\Hlep$ enters both equations with the sign of the code's own self-interaction.
The code computes $Y_p$ from the fitting formula of Ref.~\cite{Domcke:2025jiy} (Sec.~\ref{sec:bounds}). Its standard value agrees with the PRIMAT prediction $0.24709\pm0.00017$~\cite{Pitrou:2018cgg}.

With $\Hlep$ switched off, the public code reproduces the published tables of Ref.~\cite{Domcke:2025lzg} to better than $3\%$ in the final asymmetries at six randomly chosen points.
We then corrected two defects.
The public code uses the time derivative of the neutrino temperature in each step before computing it, which makes repeated runs at large asymmetry differ by up to $9\%$.
Our first implementation of the muon potential normalized the neutrino degeneracy to the comoving temperature instead of the neutrino temperature.
After both corrections, repeated runs agree to $0.5\%$ and two builds with different compilers differ by $1.6\%$, which we include as a $2\%$ reproducibility error.

\textit{Momentum-resolved code.}
To include the momentum dependence we also solve Eq.~(3) mode by mode.
A direct integration is stiff. For a sizable $\xi_{\nu_\mu}$, between $T=20$ and $8\,$MeV, the self-interaction and $\Hlep$ are much larger than the Hubble rate and implicit integrators stalled.
The large self-interaction, however, also simplifies the problem.
When it exceeds the spread of $\Omega_p$ and of the collision rates, it holds every $\rho_p$ and $\bar\rho_p$ in the eigenbasis of $D$. The corrections are of relative size $1/R$, with
\begin{equation}
R=\frac{\sqrt2\GF\min_{a\neq b}|d_a-d_b|}{\max_{p,p'}\|\Omega_p-\Omega_{p'}\|+\Delta\Gamma},
\label{eq:R}
\end{equation}
where $d_a$ are the eigenvalues of $D$ and $\Delta\Gamma$ is the spread of the collision rates over the modes.
In this synchronized limit, we write $D=Q\,{\rm diag}(d_a)Q^\dagger$, $\rho_p=Q\,{\rm diag}(f_{p,a})Q^\dagger$ and $\bar\rho_p=Q\,{\rm diag}(\bar f_{p,a})Q^\dagger$, and evolve the common eigenbasis $Q$ and the occupations $f_{p,a}$ and $\bar f_{p,a}$.
Matching the off-diagonal elements of Eq.~\eqref{eq:Dmat} in the eigenbasis of $D$ gives $\dot Q=-iH_{\rm eff}Q$ with
\begin{equation}
H_{\rm eff}=\Hlep+Q\mathcal{K}Q^\dagger,\qquad \mathcal{K}_{ab}=\frac{\int_p\Omega'_{p,ab}(u_{p,b}-u_{p,a})+iX_{ab}}{d_b-d_a}
\label{eq:Heff}
\end{equation}
for $a\neq b$, where $\Omega'=Q^\dagger\Omega Q$, $u=f+\bar f$ and $X=Q^\dagger\int_p(C_p-\bar C_p)Q$.
The occupations evolve as $\dot f_{p,a}=(Q^\dagger C_pQ)_{aa}$ and $\dot{\bar f}_{p,a}=(Q^\dagger\bar C_pQ)_{aa}$.
The self-interaction no longer appears, which removes the stiffness. For two flavors $H_{\rm eff}$ reduces to the precession axis of Eq.~(4).

We integrate $Q$ with a second-order exponential midpoint rule and the occupations with the explicit midpoint rule.
The evolution of the frame reproduces $\dot D$ computed from the full momentum-resolved right-hand side, including the self-interaction, to better than $10^{-3}$.
Both codes report $\xi_{\nu_e}^{\rm BBN}$ as defined in Sec.~\ref{sec:param}. Figure~1 of the Letter shows the instantaneous flavor asymmetries. In Fig.~1(a) the momentum-resolved points stop at $T=4\,$MeV, because below it our output spacing does not resolve the oscillation.

The synchronized limit fails where two eigenvalues of $D$ meet, because $\mathcal{K}$ in Eq.~\eqref{eq:Heff} then diverges.
On the line $\xi_{\nu_e}=\xi_{\nu_\mu}$ the muon asymmetry vanishes, and $\Hlep$ reduces to the small baryonic term. On the line $\xi_{\nu_\mu}=\xi_{\nu_\tau}$, $\Hlep$ stays large.
Apart from the valleys at $A\ge1$ (Sec.~\ref{sec:plane}), no such degeneracy occurred in the runs we report.

Collisions with $e^\pm$ and among the neutrinos damp the flavor coherence and drive the occupations toward Fermi--Dirac spectra.
The two-body processes are elastic scattering on $e^\pm$, annihilation $\nu\bar\nu\leftrightarrow e^+e^-$, and neutrino--neutrino scattering and pair processes.
They conserve each flavor lepton number. Below $T=20\,$MeV $\nu_\mu$ and $\nu_\tau$ interact in the same way, so the collision terms also commute with rotations in the $\mu\tau$ plane.
The momentum-resolved code has two versions of the collision terms. The first is approximate and fast enough for scans. The second uses the full collision integrals and tests the first. We call them the versions with approximate and with full collisions.

\textit{Approximate collision terms.}
The first version uses Maxwell--Boltzmann statistics in the matrix elements, sets $m_e\to0$, and omits $\nu\nu\to\nu\nu$ scattering.
We write $G_L={\rm diag}(g_L,g_L-1,g_L-1)$, $g_L=\frac12+s_W^2$ and $g_R=s_W^2$, with $s_W^2=\sin^2\theta_W$ the weak mixing angle, and $T_\gamma$ and $T$ for the photon and comoving neutrino temperatures.
Scattering on $e^\pm$ damps flavor coherence as
\begin{equation}
C_p\supset-k_{\rm sc}(G_L^2\rho_p+\rho_pG_L^2-2G_L\rho_pG_L),
\end{equation}
where $k_{\rm sc}$$=\frac{32}{3\pi^3}\GF^2T_\gamma^4p$.
Annihilation into $e^+e^-$ gives
\begin{equation}
C_p\supset k_{\rm ann}\Big[(G_L^2+g_R^2)f_{\rm eq}(p)\frac{T_\gamma^4}{T^4}-\frac12\{\rho_p,G_L\langle\bar\rho\rangle G_L\}-\frac{g_R^2}{2}\{\rho_p,\langle\bar\rho\rangle\}\Big],
\end{equation}
with $k_{\rm ann}$$=8\GF^2T^4p/3\pi^3$, $f_{\rm eq}$ the Fermi--Dirac distribution at $T_\gamma$ without chemical potential, and $\langle\bar\rho\rangle$ the momentum-weighted average of $\bar\rho_p$, normalized to one in equilibrium.
Pair conversion $\nu\bar\nu\leftrightarrow\nu\bar\nu$ gives
\begin{equation}
C_p\supset k_{\rm nn}[{\rm Tr}(\rho_p\langle\bar\rho\rangle+\langle\bar\rho\rangle\rho_p)\openone-3(\rho_p\langle\bar\rho\rangle+\langle\bar\rho\rangle\rho_p)],
\end{equation}
with $k_{\rm nn}$$=2\GF^2T^4p/3\pi^3$ and $\openone$ the unit matrix.
Elastic scattering also relaxes the diagonal occupations toward Fermi--Dirac spectra at $T_\gamma$ with the current flavor numbers, at the total scattering rate $64(g_{L\alpha}^2+g_R^2)\GF^2T_\gamma^4p/3\pi^3$, where $g_{L\alpha}$ are the diagonal entries of $G_L$.
The momenta lie on the nodes of generalized Gauss--Laguerre quadrature in $p/T$.
For the $A=1$ benchmark, halving the step from $0.005$ to $0.0025$ changes $\xi_{\nu_e}$ at $T=0.8\,$MeV by $2\%$, and $4$, $8$ and $16$ momentum modes agree to $1\%$.

The collision rates are fixed by the Standard Model. To test that the effect of $\Hlep$ does not come from our collision terms, whose uncertainty is about $14\%$ in $\xi_{\nu_e}^{\rm BBN}$ (Sec.~\ref{sec:bounds}), we halved and doubled all rates. For the $A=1$ benchmark along $L_\mu-L_\tau$, halved, nominal and doubled rates give $|\xi_{\nu_e}^{\rm BBN}|=0.24$, $0.17$ and $0.10$ with approximate collisions and $0.25$, $0.18$ and $0.11$ with full collisions. Without $\Hlep$ the version with full collisions gives $\xi_{\nu_e}^{\rm BBN}=+0.011$, $-0.014$ and $-0.011$, so the ratio stays between $10$ and $23$.

\textit{Full collision integrals.}
The second version removes these approximations. It uses the full two-body collision integrals of Refs.~\cite{Froustey:2021azz,Froustey:2024mgf}, with Fermi--Dirac statistics, the electron mass, all neutrino--neutrino processes, the complete flavor structure and $\bar\rho\neq\rho$.
The momenta lie on $40$ uniformly spaced points, on which the neutrino--neutrino terms conserve lepton numbers and energy exactly.
At zero asymmetry, without oscillations, QED corrections and muons in both codes, it reproduces $N_{\rm eff}$ of FortEPiaNO~\cite{Bennett:2020zkv} to $10^{-4}$.
At $\xi_{\nu_\mu}=0.071$ with $\Hlep$, $20$ or $60$ instead of $40$ momenta and halved steps change $\xi_{\nu_e}^{\rm BBN}$ by less than $3\%$.
Without $\Hlep$ it reproduces the final $\xi_{\nu_e}$ of the full neutrino transport of Ref.~\cite{Froustey:2024mgf}, released as public grids, to $2\%$ at $|\xi_{\nu_\mu}|=0.28$, $0.55$ and $0.80$ along $L_\mu-L_\tau$ for $\xi_{\nu_\mu}>0$, and to $5\%$ at $0.14$, $0.42$ and $0.68$ for $\xi_{\nu_\mu}<0$. The momentum-averaged code lies $19$ to $31\%$ above the full transport there.

\begin{figure}[t]
\includegraphics[width=0.6\columnwidth]{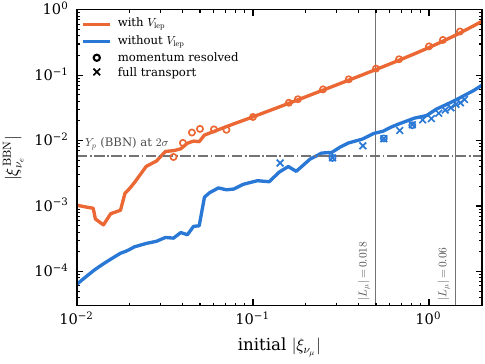}
\caption{$\xi_{\nu_e}^{\rm BBN}$ along $L_\mu-L_\tau$, as Fig.~2(a) of the Letter but for initial $\xi_{\nu_\mu}=-\xi_{\nu_\tau}>0$ and $\xi_{\nu_e}=0$, for normal mass ordering, with $\Hlep$ (orange) and without (blue). Lines show the momentum-averaged solution. Circles show our momentum-resolved solution with full collision integrals, for $|\xi_{\nu_\mu}|\ge0.035$ with $\Hlep$ and $0.28\le|\xi_{\nu_\mu}|\le0.81$ without. Crosses show the full neutrino transport of Ref.~\cite{Froustey:2024mgf} without $\Hlep$. The degeneracy $\xi_{\nu_e}^{\rm BBN}$ is negative, except with $\Hlep$ for $|\xi_{\nu_\mu}|<0.018$. The gray line marks where $Y_p$ alone leaves its $2\sigma$ range for $\xi_{\nu_e}^{\rm BBN}<0$, without the theory error on $\xi_{\nu_e}^{\rm BBN}$ and without $N_{\rm eff}$. Vertical lines mark $|L_\mu|=0.06$ and $0.018$.}
\label{fig:smline}
\end{figure}

\textit{Comparison of the codes.} Table~\ref{tab:bench} compares the two codes at benchmark points. Along $L_\mu-L_\tau$, Fig.~2(a) of the Letter compares them for $\xi_{\nu_\mu}<0$ and Fig.~\ref{fig:smline} for $\xi_{\nu_\mu}>0$.
With $\Hlep$ and normal ordering the momentum-averaged code and the momentum-resolved code with approximate collisions agree to within $7\%$ for $|\xi_{\nu_\mu}|\ge0.04$ along $L_\mu-L_\tau$, and to within $3\%$ for the tauphobic benchmarks in both orderings.
The momentum-resolved code with full collisions agrees with the momentum-averaged code to within $10\%$ for $\xi_{\nu_\mu}<0$ and $0.055\le|\xi_{\nu_\mu}|\le1.5$, and to within $25\%$ at $0.04$ and $0.045$. Below $0.04$ the two differ by a factor of two or more.
For $\xi_{\nu_\mu}>0$ they agree to within $5\%$ for $0.1\le|\xi_{\nu_\mu}|\le1.5$, but differ by up to $50\%$ at $0.035$ to $0.07$, where the crossing is partly non-adiabatic.
For inverted ordering individual values differ by up to $0.05$, because $\xi_{\nu_e}^{\rm BBN}$ depends on the phase of collective oscillations when collisions freeze out.

\begin{table}[t]
\caption{$\xi_{\nu_e}^{\rm BBN}$ without and with $\Hlep$ in the momentum-averaged code and the momentum-resolved code with approximate collisions. The $L_\mu-L_\tau$ rows have $\xi_{\nu_\mu}>0$ and amplitude $A$. Tauphobic means $\xi_{\nu_e}=-\xi_{\nu_\mu}=0.23$ and $\xi_{\nu_\tau}=0$.}
\label{tab:bench}
\begin{tabular}{llcccc}
\hline\hline
 & & \multicolumn{2}{c}{momentum averaged} & \multicolumn{2}{c}{momentum resolved}\\
order & direction & without & with & without & with\\
\hline
NH & $L_\mu-L_\tau$, $A=0.06$ & $-0.0004$ & $-0.0091$ & $-0.0019$ & $-0.0095$\\
NH & $L_\mu-L_\tau$, $A=0.15$ & $-0.0024$ & $-0.0245$ & $-0.0036$ & $-0.0245$\\
NH & $L_\mu-L_\tau$, $A=1$ & $-0.0155$ & $-0.170$ & $-0.0139$ & $-0.174$\\
NH & tauphobic & $-0.0197$ & $+0.0591$ & $-0.0170$ & $+0.0598$\\
IH & $L_\mu-L_\tau$, $A=0.1$ & $-0.0003$ & $+0.0122$ & $--$ & $+0.0115$\\
IH & $L_\mu-L_\tau$, $A=1$ & $-0.0057$ & $+0.0526$ & $-0.0058$ & $+0.0529$\\
IH & tauphobic & $+0.0435$ & $-0.0218$ & $+0.0211$ & $-0.0213$\\
\hline\hline
\end{tabular}
\end{table}

Without $\Hlep$ both versions of the momentum-resolved code fail for $|\xi_{\nu_\mu}|\gtrsim0.9$. At $|\xi_{\nu_\mu}|=0.97$ and $1.23$ with $\xi_{\nu_\mu}>0$ the version with full collisions gives $\xi_{\nu_e}^{\rm BBN}=+0.082$ and $+0.119$, against $-0.021$ and $-0.029$ from the full transport, and the version with approximate collisions fails in the same way. The failure therefore comes from the synchronized limit and not from the collision terms. It occurs although $R$ is $150$ to $200$ there. A likely cause, which we have not tested, is that along $L_\mu-L_\tau$ the vacuum part of $\mathcal{K}_{\mu\tau}$ in Eq.~\eqref{eq:Heff} vanishes, because $u_{p,\mu}=u_{p,\tau}$, so the terms of order $1/R$ that the synchronized limit drops drive the $\mu\tau$ dynamics. Without $\Hlep$ the $\mu$ and $\tau$ asymmetries swap through a non-adiabatic transition, while with $\Hlep$ the flavors stay locked until the crossing.
The Letter therefore uses the full transport as the reference without $\Hlep$ in Fig.~2(a).

\textit{Tests of the synchronized limit.}
The momentum-resolved code assumes that all modes share the eigenbasis of $D$. We test this in three ways, with a stability analysis, with a solution in which each mode follows its own eigenbasis, and with a solution that evolves every mode as a full matrix.

First, we linearized the momentum-resolved equations about $\rho_p=Q\,{\rm diag}(f_{p,a})Q^\dagger$. For each pair of eigenvectors of $D$ this gives a dispersion problem for the coherences between them. We drop collisions and the couplings between different pairs, which are of order $1/R$, and evaluate it on snapshots of the code with full collisions. No mode grows above $T=2\,$MeV in the five cases we tested, which include the locking and the crossing. Near the bound for $\xi_{\nu_\mu}<0$ a mode starts to grow between $T=1.6$ and $1.5\,$MeV, at $300$ times the Hubble rate at $T=1.5\,$MeV. No mode grows near the bound for $\xi_{\nu_\mu}>0$, at $\xi_{\nu_\mu}=-0.68$ or without $\Hlep$ at $|\xi_{\nu_\mu}|=0.97$. The analysis does not fix the onset of the instability and does not prove stability elsewhere.

Second, we solved the equations in the adiabatic approximation of Ref.~\cite{Froustey:2021azz}, the adiabatic transfer of averaged oscillations (ATAO), in its ATAO-(J$\pm$V) form, with the collision integrals and momentum grid of the momentum-resolved code. It evolves the asymmetry matrix directly and keeps each mode diagonal in the eigenbasis of its own Hamiltonian. Without $\Hlep$ it reproduces the full transport to $2.5\%$ or better at $|\xi_{\nu_\mu}|=0.28$, $0.55$ and $0.80$.
With $\Hlep$ it gives $\xi_{\nu_e}^{\rm BBN}=0.0190$ and $-0.0051$ for $\xi_{\nu_\mu}=-0.065$ and $+0.035$, against $0.0190$ and $-0.0056$ from the momentum-resolved code. With its values the bound for $\xi_{\nu_\mu}>0$ becomes $0.036$ to $0.038$, against $0.035$ to $0.037$ from the momentum-resolved code alone.
This tests the assumption of a common basis for all momenta, but not the instability, because both codes keep every mode diagonal in a slowly varying basis.

Third, we evolved every momentum mode as a full $3\times3$ matrix with its own kinetic equation, which we call the full-matrix solution. Below $T=2\,$MeV we start from the state of the momentum-resolved code at $T=2\,$MeV and keep its Hamiltonian, momentum grid and collision integrals, which now act on the full matrices. We follow the rotation by the self-interaction, about $10^9$ radians per Hubble time, with exponential midpoint steps of at most $0.3$ radians, and $0.15$ radians below $T=1.2\,$MeV.
Near the bound for $\xi_{\nu_\mu}<0$ the off-diagonal components of the modes in the eigenbasis of $D$ stay at a few times $10^{-6}$ of the diagonal ones down to $T=1.51\,$MeV. They then grow at about $250$ times the Hubble rate and saturate between $3$ and $6\times10^{-5}$. The full-matrix solution gives $\xi_{\nu_e}^{\rm BBN}=0.01898$, against $0.01901$ from the momentum-resolved code and $0.0190$ from ATAO. At $T=0.8\,$MeV it gives $0.0175$, against $0.0166$ from the momentum-resolved code and $0.0176$ from ATAO. Halving the step from $T=1\,$MeV on raises $\xi_{\nu_e}$ by $2.1\%$ at $T=0.84\,$MeV. 
These differences are smaller than the change of $13\%$ between the read-outs at $T=1.5$ and $0.8\,$MeV, which the range of the bound already includes. Near the bound for $\xi_{\nu_\mu}>0$ our steps are not accurate enough, and we do not use it there.

The full-matrix solution also follows the crossing, from $T=6$ to $3\,$MeV. We start from the state of the momentum-resolved code at $T=6\,$MeV, where the flavors are still locked, and keep its Hamiltonian, momentum grid and collision integrals. With the physical self-interaction these equations are too stiff, so we multiply it by $\kappa=10^{-3}$, $3\times10^{-3}$ and $10^{-2}$.
Near the bound for $\xi_{\nu_\mu}<0$, at $T=3\,$MeV this solution gives $40\%$, $85\%$ and $104\%$ of $\xi_{\nu_e}$ from the momentum-resolved code. At $\kappa=10^{-2}$ it agrees to within $4\%$ between $T=4.5$ and $3\,$MeV and is $13\%$ lower at $T=5\,$MeV, during the crossing. The off-diagonal components of the modes in the eigenbasis of $D$ stay below $3\times10^{-5}$, and at $\kappa=10^{-3}$ halving the step changes $\xi_{\nu_e}$ by $1\%$.
Near the bound for $\xi_{\nu_\mu}>0$, where $|\xi_{\nu_e}|$ stays below $0.006$ and the crossing is partly non-adiabatic, the results do not converge with $\kappa$, and we do not use them.

\textit{Tests of the potential.}
First, we let the muon asymmetry relax toward chemical equilibrium at the rate set by muon decay, $\Gamma_\mu K_1(m_\mu/T)/K_2(m_\mu/T)$, instead of imposing equilibrium. This changes $\xi_{\nu_e}^{\rm BBN}$ along $L_\mu-L_\tau$ by at most $0.1\%$, and by at most $4\%$ for a rate one hundred times smaller.
The potential has no free parameter, because chemical equilibrium and charge neutrality fix it. As a stress test we scaled it by $0.3$ and $3$. This changes $\xi_{\nu_e}^{\rm BBN}$ by at most $7\%$ at $A=1$ and $6\%$ at $A=0.1$, because only the time at which the potential falls below the vacuum term matters. Near the bound for $\xi_{\nu_\mu}>0$, at $A=0.05$, where the crossing is least adiabatic (Sec.~\ref{sec:crossing}), it changes $\xi_{\nu_e}^{\rm BBN}$ by $-55\%$ and $+16\%$.
The same test bounds the effect of reading the degeneracies in Eq.~(1) off the diagonal of $D$, although near the crossing the spectra are not exactly thermal.

Second, we wrote a collisionless calculation from Eqs.~(1), (2), \eqref{eq:dnmu} and \eqref{eq:Dmat} alone, which shares only physical constants with the other codes. It takes every mode to be a Fermi--Dirac distribution in the eigenbasis of $D$, with degeneracies given by the eigenvalues of a matrix $\Xi$. With $\Omega_p=UM^2U^\dagger/2p+p\,v_c$, where $p\,v_c$ is the charged-current thermal potential of $e^\pm$~\cite{Notzold:1987ik}, and with $\hat D=D/T^3=\Xi/6+\Xi^3/6\pi^2$, one finds
\begin{equation}
\frac{d\hat D}{d\ln a}=-\frac{i}{H}\left\{\frac{[UM^2U^\dagger,\Xi^2]}{8\pi^2T}+T\left[v_c,\frac{\Xi^2}{4}+\frac{\Xi^4}{8\pi^2}\right]+[\Hlep,\hat D]\right\}.
\label{eq:X12}
\end{equation}
The eigenvalues of $\hat D$ are conserved, so $\Xi$ follows from $\hat D$. With the collision rates of the synchronized code set to zero, the two agree on $\xi_{\nu_e}^{\rm BBN}$ to $0.07\%$ at the bound for $\xi_{\nu_\mu}<0$ and $0.5\%$ at the bound for $\xi_{\nu_\mu}>0$. Through the crossing at $|\xi_{\nu_\mu}|=0.35$ they agree to $1\%$.

\section{The level crossing}\label{sec:crossing}
This section describes the crossing of Fig.~1(b) of the Letter in terms of the levels of the effective Hamiltonian $H_{\rm eff}$ of Eq.~\eqref{eq:Heff}.

\begin{figure}[t]
\includegraphics[width=0.55\columnwidth]{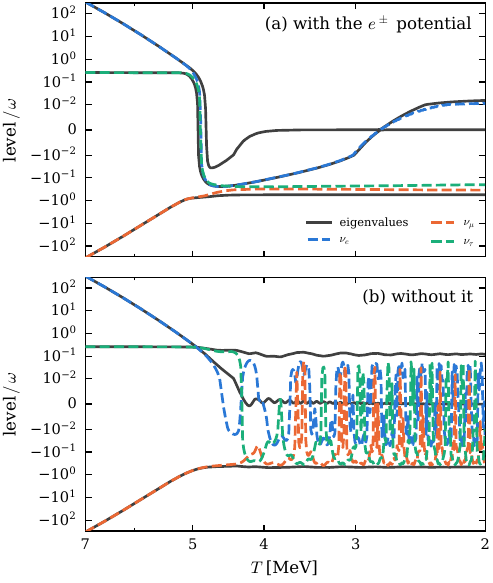}
\caption{Levels at the crossing for $A=1$ along $L_\mu-L_\tau$, initial $\xi_{\nu_\alpha}=(0,-0.68,0.68)$, normal ordering, with $\Hlep$, (a) with and (b) without the thermal potential of $e^\pm$, from the collisionless calculation of Sec.~\ref{sec:codes}. Solid lines are the eigenvalues of the effective Hamiltonian $H_{\rm eff}$ of Eq.~\eqref{eq:Heff}, which rotates the eigenbasis of the asymmetry matrix. Dashed lines are its flavor-diagonal entries, the levels of the flavor states without mixing. Both are in units of $\omega=\Delta m^2_{31}/2\langle E\rangle$. The diagonal of $H_{\rm eff}$ in the eigenbasis of the asymmetry matrix, which does not enter the evolution, is fixed by continuity (Sec.~\ref{sec:crossing}).}
\label{fig:levels}
\end{figure}

\textit{Pair weights.} In the eigenbasis of $D$ the vacuum term couples the eigenstates $a$ and $b$ through $\int_p\Omega'_{p,ab}(u_{p,b}-u_{p,a})/(d_b-d_a)$, as in Eq.~\eqref{eq:Heff}. Here $d_a$ are the eigenvalues of $D$, and $\Omega'_p$ is $\Omega_p$ in its eigenbasis. In that basis $u_{p,a}=f_{p,a}+\bar f_{p,a}$ are the occupations of mode $p$, and $\xi_a$ are the degeneracies of the eigenstates (Sec.~\ref{sec:codes}). Relative to $\omega$ of Sec.~\ref{sec:eqs}, the weight of the pair is $w_{ab}=\langle E\rangle\int_p p^{-1}(u_{p,b}-u_{p,a})/(d_b-d_a)$. For Fermi--Dirac spectra and small degeneracies $w_{ab}=(9.45/2\pi^2)(\xi_a+\xi_b)\simeq0.48(\xi_a+\xi_b)$. The two-flavor weight ${\bm S} \cdot {\bm D}/D^2$ of the Letter, which takes the vacuum term at one mean momentum, is $0.42(\xi_a+\xi_b)$. Along $L_\mu-L_\tau$ the $\nu_\mu\nu_\tau$ pair has zero weight, and the $\nu_e\nu_\mu$ and $\nu_e\nu_\tau$ pairs have weights of opposite sign.

\textit{Levels.} We call the eigenvalues of $H_{\rm eff}$ in Eq.~\eqref{eq:Heff} the levels of the crossing. Each belongs to one eigenstate of the asymmetry matrix, which follows it when the evolution is adiabatic. The flavor-diagonal entries of $H_{\rm eff}$ are the levels of the flavor states without mixing. Only the entries of $\mathcal{K}$ with $a\neq b$ enter the evolution. A diagonal part only changes the phases of the eigenvectors, but it does shift the levels, so we fix it by continuity, $\mathcal{K}_{aa}$$=\int_p\Omega'_{p,aa}\,\partial u_{p,a}/\partial d_a$, the limit of $\mathcal{K}_{ab}$ for $d_b\to d_a$. The levels therefore illustrate the crossing. The measurable results are whether the asymmetry follows the eigenstates and how much it transfers (Table~\ref{tab:adiab}). When $\Hlep$ fades, the $\nu_e$ level crosses the $\nu_\tau$ level (Fig.~\ref{fig:levels}), which takes the $\nu_\tau$ asymmetry to $\nu_1$ and the $\nu_\mu$ asymmetry to $\nu_3$.

\textit{Adiabaticity.} The crossing is adiabatic when the levels move slowly compared with the precession across the gap between them, so that the asymmetry follows the eigenstates of $H_{\rm eff}$. We measure it by the transfer it produces. Without collisions an adiabatic crossing would put $|U_{e1}|^2-|U_{e3}|^2=0.656$ of the initial $\nu_\tau$ asymmetry into $\nu_e$. Table~\ref{tab:adiab} gives the fraction of this transfer reached by the collisionless calculation of Sec.~\ref{sec:codes}. It is $0.64$ at the bound for $\xi_{\nu_\mu}<0$, $0.32$ at the bound for $\xi_{\nu_\mu}>0$ and above $0.9$ at $A=1$. It grows with $|\xi_{\nu_\mu}|$ because the pair weights, and with them the gap at the crossing, grow with the degeneracies, while $\Hlep$ falls at a rate set by $m_\mu/T$. The two bounds differ mainly because they lie at different $|\xi_{\nu_\mu}|$. With collisions $\xi_{\nu_e}^{\rm BBN}$ depends less on adiabaticity. The ratio $\xi_{\nu_e}^{\rm BBN}/|\xi_{\nu_\mu}|$ is $0.29$ at the bound for $\xi_{\nu_\mu}<0$ and $0.25$ at $A=1$.

\begin{table}[t]
\caption{Fraction of the adiabatic transfer, $0.656$ of the initial $\nu_\tau$ asymmetry, that reaches $\nu_e$ at $T=1.5\,$MeV without collisions, along $L_\mu-L_\tau$ with $\Hlep$ for normal ordering. Asymmetries are compared as $\xi+\xi^3/\pi^2$. The last column is $A=1$.}
\label{tab:adiab}
\begin{tabular}{lccccccc}
\hline\hline
$|\xi_{\nu_\mu}|$ & $0.035$ & $0.04$ & $0.045$ & $0.065$ & $0.1$ & $0.35$ & $0.68$\\
\hline
$\xi_{\nu_\mu}<0$ & -- & $0.34$ & $0.49$ & $0.64$ & $0.78$ & $0.82$ & $0.94$\\
$\xi_{\nu_\mu}>0$ & $0.32$ & $0.40$ & $0.41$ & -- & $0.69$ & $0.97$ & $0.91$\\
\hline\hline
\end{tabular}
\end{table}

\textit{Role of the thermal potential of $e^\pm$.} Near $T=5\,$MeV the thermal potential of $e^\pm$, weighted like the vacuum term, is as large as the vacuum term. This is the electron-driven transition of Ref.~\cite{Domcke:2025lzg}. Once $\Hlep$ has faded, it pushes the $\nu_e$ level well below the $\nu_\tau$ level, and the asymmetry follows the eigenstates (Fig.~\ref{fig:levels}a).
We switched it off with $\Hlep$ on, in the momentum-resolved code with full collisions, where $(\varepsilon+P)_e$ enters only $\Omega_p$, and in the collisionless calculation. Without it the levels nearly meet and the asymmetry oscillates between them (Fig.~\ref{fig:levels}b).
With collisions $\xi_{\nu_e}^{\rm BBN}$ then falls by factors of $0.28$ at the bound for $\xi_{\nu_\mu}<0$ and $0.21$ at $A=1$, and nearly vanishes at the bound for $\xi_{\nu_\mu}>0$. Without collisions the fraction of the adiabatic transfer falls from $0.64$ to about zero at the bound for $\xi_{\nu_\mu}<0$ and from $0.94$ to $0.31$ at $A=1$. The crossing, where $|D_{ee}|$ reaches half its maximum, moves from $T=4.9$ to $4.4\,$MeV.
The transfer into $\nu_e$ therefore needs both $\Hlep$, which keeps the flavors locked past the muon-driven transition, and the $e^\pm$ potential, which takes over from it at about the same temperature.

\section{Helium bounds}\label{sec:bounds}

\textit{Helium abundance and $N_{\rm eff}$.} The momentum-averaged code evolves the neutrino and photon temperatures together with the asymmetries. After $e^\pm$ annihilation it gives $N_{\rm eff}=\frac87\big(\frac{11}{4}\big)^{4/3}\varepsilon_\nu/\varepsilon_\gamma$ from the neutrino and photon energy densities $\varepsilon_\nu$ and $\varepsilon_\gamma$, normalized to $3.044$ without asymmetries, and $\Delta N_{\rm eff}=N_{\rm eff}-3.044$~\cite{Domcke:2025lzg}. For Fermi--Dirac spectra with degeneracies $\xi_{\nu_\alpha}$ the asymmetries add $\Delta N_{\rm eff}=\sum_\alpha[\frac{30}{7}(\xi_{\nu_\alpha}/\pi)^2+\frac{15}{7}(\xi_{\nu_\alpha}/\pi)^4]$. The code computes $Y_p$ from the fit in Eq.~(2.22a) of Ref.~\cite{Domcke:2025jiy} to their BBN network at fixed $\eta_B$,
\begin{equation}
Y_p \simeq 0.2470\,\exp\!\big(-0.961\,\xi_{\nu_e}^{\rm BBN}-0.185\,(\xi_{\nu_e}^{\rm BBN})^2+0.053\,\Delta N_{\rm eff}-0.682\,\delta T+\dots\big),
\label{eq:Ypfit}
\end{equation}
where the dots stand for smaller and mixed terms. Here $\xi_{\nu_e}^{\rm BBN}=6\Delta n_{\nu_e}/T_{\nu_e}^3$ at $T=1.5\,$MeV, with $T_{\nu_e}$ from the $\nu_e$ energy density, and $\delta T=T_{\nu_e}/T_\gamma-(4/11)^{1/3}$ after $e^\pm$ annihilation. This $\xi_{\nu_e}^{\rm BBN}$ differs from ours by less than $0.1\%$.
The leading term reflects the neutron-to-proton ratio, which before weak freeze-out follows $\exp(-\delta m/T-\xi_{\nu_e})$, with $\delta m$ the neutron--proton mass difference. To first order
\begin{equation}
Y_p=0.2470-0.237\,\xi_{\nu_e}^{\rm BBN}+0.013\,\Delta N_{\rm eff}.
\label{eq:Yplin}
\end{equation}
For the synchronized code we use Eq.~\eqref{eq:Yplin}, with $\Delta N_{\rm eff}$ from the momentum-averaged code. With Eq.~\eqref{eq:Yplin} the momentum-averaged code gives bounds within $4\%$ of those from Eq.~\eqref{eq:Ypfit}.

\textit{Likelihood.} We constrain the primordial asymmetries with
\begin{equation}
\chi^2=\frac{(Y_p-Y_p^{\rm obs})^2}{\sigma_{\rm obs}^2+\sigma_{\rm th}^2+(0.237\,\sigma_\xi)^2}+\frac{(N_{\rm eff}-N_{\rm eff}^{\rm obs})^2}{\sigma_N^2}.
\label{eq:chi2}
\end{equation}
Here $Y_p$ and $N_{\rm eff}$ are the computed values and the superscript obs marks the measurements.
For helium we use the value of the Large Binocular Telescope (LBT) $Y_p$ project, $Y_p^{\rm obs}=0.2458\pm0.0013$~\cite{Aver:2026dxv} and the BBN theory error $\sigma_{\rm th}=0.00017$~\cite{Pitrou:2018cgg}.
For $N_{\rm eff}$ we use $3.01\pm0.11$ from CMB and BAO data without abundance information~\cite{Goldstein:2026iuu}, so that helium is not counted twice. That fit predicts $Y_p$ in the CMB from standard BBN. Near our bounds with $\Hlep$ the asymmetries add $\Delta N_{\rm eff}<0.01$, so the $N_{\rm eff}$ term matters only for the bounds without $\Hlep$. The uncertainty of $\eta_B$ moves $Y_p$ by less than $10^{-4}$.
The factor $0.237=|\partial Y_p/\partial\xi_{\nu_e}^{\rm BBN}|$ converts a theory error $\sigma_\xi$ on $\xi_{\nu_e}^{\rm BBN}$ into helium.
We take $\sigma_\xi$ as the quadrature sum of three parts. The first is the difference between the momentum-averaged code and the synchronized code with approximate collisions at $T=1.5\,$MeV.
The second, $14\%$ of $\xi_{\nu_e}^{\rm BBN}$, is the largest shift of $\xi_{\nu_e}^{\rm BBN}$ at $A=0.1$ and $1$ when all collision rates are scaled by $0.8$ or $1.25$. This range is meant to cover the Fermi--Dirac and electron-mass corrections that the approximate collision terms neglect.
The third is the reproducibility error, $2\%$ of $\xi_{\nu_e}^{\rm BBN}$ with $\Hlep$ and $15\%$ without.
Along a line in flavor space we quote $95\%$ confidence bounds from $\Delta\chi^2=3.84$ relative to the minimum over the whole line, both signs included. In the plane we use $\Delta\chi^2=5.99$.
Table~\ref{tab:bounds} lists the bounds along $L_\mu-L_\tau$ for three treatments. The second assigns the overall momentum-averaging error $\sigma_\xi=0.016$ of Ref.~\cite{Domcke:2025lzg} to every point. The third uses the criteria of Ref.~\cite{Domcke:2025lzg}, $-0.040<\xi_{\nu_e}^{\rm BBN}<0.042$ and $\Delta N_{\rm eff}<0.23$.
In every treatment the bound tightens by a factor between about $6$ and $13$. The expected bounds, computed with data equal to the standard prediction, tighten by factors of $12$ and $19$.

\begin{table}[t]
\caption{$95\%$ confidence upper bounds on the initial $|\xi_{\nu_\mu}|$ along $L_\mu-L_\tau$ for normal ordering, without and with $\Hlep$, from the momentum-averaged code.}
\label{tab:bounds}
\begin{tabular}{lcccc}
\hline\hline
 & \multicolumn{2}{c}{$\xi_{\nu_\mu}>0$} & \multicolumn{2}{c}{$\xi_{\nu_\mu}<0$}\\
treatment & without & with & without & with\\
\hline
Eq.~\eqref{eq:chi2}, $\sigma_\xi$ as in text & $0.29$ & $0.035$ & $0.84$ & $0.065$\\
Eq.~\eqref{eq:chi2}, $\sigma_\xi=0.016$ & $0.70$ & $0.12$ & $0.86$ & $0.15$\\
criteria of Ref.~\cite{Domcke:2025lzg} & $1.0$ & $0.17$ & $1.0$ & $0.17$\\
expected, $\sigma_\xi$ as in text & $0.62$ & $0.050$ & $0.82$ & $0.043$\\
\hline\hline
\end{tabular}
\end{table}
\textit{Robustness.} The code difference is not a calibrated uncertainty. We also treated the theory error as a normalized Gaussian, which adds $\ln\sigma_Y^2$ to $-2\ln\mathcal L$, with $\sigma_Y^2$ the denominator of the helium term in Eq.~\eqref{eq:chi2}. With this treatment or with each code alone, the bounds with $\Hlep$ are $|\xi_{\nu_\mu}|<0.061$ to $0.065$ for $\xi_{\nu_\mu}<0$ and $0.031$ to $0.037$ for $\xi_{\nu_\mu}>0$.
With the full collision integrals they are $0.058$ to $0.066$ and $0.035$ to $0.037$, for the read-out at $T=1.5$ or $0.8\,$MeV and a theory error of zero or of $14\%$ and $2\%$ in quadrature.
Between these read-outs the synchronization margin $R$ of Eq.~\eqref{eq:R} falls from $11$ to $0.7$ at $\xi_{\nu_\mu}=-0.065$, as collisions bring two eigenvalues close.
With the full transport of Ref.~\cite{Froustey:2024mgf} and its helium abundances, the bounds without $\Hlep$ are $0.27$ for $\xi_{\nu_\mu}>0$ and $0.64$ for $\xi_{\nu_\mu}<0$, and the tightening factors become about $8$ and $10$.
With $\theta_{23}$ in the lower octant ($42.4^\circ$) or with $\delta_{\rm CP}=\pi/2$ or $\pi$, $\xi_{\nu_e}^{\rm BBN}$ at the bound for $\xi_{\nu_\mu}<0$ changes by at most $9\%$. At the bound for $\xi_{\nu_\mu}>0$ (Sec.~\ref{sec:crossing}) it changes by up to $31\%$ in the momentum-resolved code with full collisions and $7\%$ in the momentum-averaged code.
We have not computed the effect of the non-thermal $\nu_e$ spectrum on $Y_p$.

\section{The plane of zero total lepton number}\label{sec:plane}

We scanned the plane with the momentum-averaged code in steps of $10^\circ$ in $\phi$ for normal ordering. In the plane the theory error $\sigma_\xi$ uses $10\%$ of $\xi_{\nu_e}^{\rm BBN}$ in place of the code difference.

\begin{figure}[t]
\includegraphics[width=0.65\columnwidth]{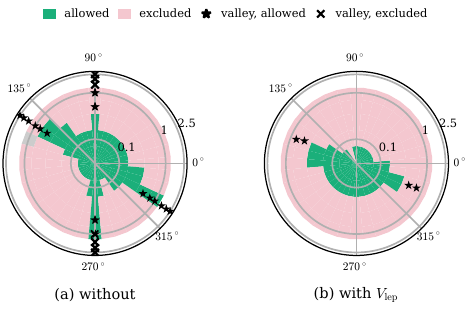}
\caption{Directions in the plane of zero total lepton number allowed at $95\%$ confidence for normal ordering, (a) without and (b) with $\Hlep$. The radius is the amplitude $A$ on a logarithmic scale and the angle is $\phi$. Shading shows a grid with steps of $10^\circ$, and gray marks grid points where the run did not converge. Circles mark $A=0.1$, $1$ and $2.5$. Stars and crosses mark allowed and excluded zeros of $\xi_{\nu_e}^{\rm BBN}$, located by bisection to $0.4^\circ$, for $A\le0.75$ with $\Hlep$ and for all amplitudes without it. $L_\mu-L_\tau$ corresponds to $\phi=90^\circ$ and $270^\circ$, tauphobic asymmetries to $135^\circ$ and $315^\circ$.}
\label{fig:plane}
\end{figure}

\textit{Generic directions.} Reference~\cite{Domcke:2025lzg} found $|\xi_{\nu_e}^{\rm BBN}|\le0.171A$ for normal ordering, which we reproduce without $\Hlep$. With $\Hlep$ the largest ratio is $0.21$, so the minimal washout factor falls from about $6$ to about $5$.
The sign of $\xi_{\nu_e}^{\rm BBN}$ reverses in $80$ to $90\%$ of all directions at every amplitude we computed. The momentum-resolved code with approximate collisions confirms the reversal at five generic directions at $A=0.25$, where the two codes agree to within $5\%$.
Figure~\ref{fig:plane} shows the allowed directions.

\begin{table}[t]
\caption{Zeros of $\xi_{\nu_e}^{\rm BBN}$ in the plane of zero total lepton number for normal ordering, located by bisection to $0.4^\circ$, and whether they are allowed at $95\%$ confidence. Each entry gives $\phi$ and, in parentheses, $N_{\rm eff}$ at the sign change of $\xi_{\nu_e}^{\rm BBN}$. Not determined means that $\xi_{\nu_e}^{\rm BBN}$ varies erratically within a degree, so that the bisection cannot locate a single sign change.}
\label{tab:valleys}
\begin{tabular}{lll}
\hline\hline
$A$ & without $\Hlep$ & with $\Hlep$\\
\hline
$0.5$ & $90^\circ$ (3.07), $148^\circ$ (3.06), $270^\circ$ (3.07), $328^\circ$ (3.06) & $157^\circ$ (3.05), $337^\circ$ (3.05)\\
$0.75$ & $148^\circ$ (3.07), $328^\circ$ (3.07) & $158^\circ$ (3.05), $338^\circ$ (3.05)\\
$1$ & $90^\circ$ (3.15), $148^\circ$ (3.09), $270^\circ$ (3.15), $328^\circ$ (3.09) & not determined\\
$1.5$ & $90^\circ$ (3.26)$^\times$, $147^\circ$ (3.12), $270^\circ$ (3.26)$^\times$, $327^\circ$ (3.12) & not determined\\
$2$ & $90^\circ$ (3.39)$^\times$, $147^\circ$ (3.16), $270^\circ$ (3.39)$^\times$, $327^\circ$ (3.16) & not determined\\
$2.5$ & $90^\circ$ (3.52)$^\times$, $147^\circ$ (3.19), $270^\circ$ (3.51)$^\times$, $327^\circ$ (3.19) & not determined\\
\hline\hline
\multicolumn{3}{l}{\footnotesize $^\times$ excluded at $95\%$ confidence.}
\end{tabular}
\end{table}

\textit{Near $L_\mu-L_\tau$.} Without $\Hlep$, $\xi_{\nu_e}^{\rm BBN}\simeq0.017A$ along $L_\mu-L_\tau$ and changes sign within a fraction of a degree at some amplitudes, the non-adiabatic knife edge of Ref.~\cite{Domcke:2025lzg}. This small value needs $\xi_{\nu_e}=0$ exactly. Ten degrees away in $\phi$ it is three to ten times larger and of opposite sign, while with $\Hlep$ it changes by about $10\%$, and the valley along $L_\mu-L_\tau$ disappears.

\textit{Valleys.} Narrow valleys, where $\xi_{\nu_e}^{\rm BBN}$ passes through zero, allow large asymmetry amplitudes~\cite{Domcke:2025lzg}. We located them by bisection in $\phi$ at fixed amplitude (Table~\ref{tab:valleys}).
Without $\Hlep$ there are smooth valleys near $\phi=148^\circ$ and $328^\circ$, allowed up to $A=2.5$, the edge of our scan, and limited only by $N_{\rm eff}$.
With $\Hlep$ the valleys move to $\phi\simeq157^\circ$ and $337^\circ$, and at $A=0.5$ and $0.75$ they are smooth zeros in allowed directions.
At $A\ge1$ with $\Hlep$, $\xi_{\nu_e}^{\rm BBN}$ near the shifted valleys varies erratically on scales below one degree, for example between $-0.05$ and $+0.04$ within $2^\circ$ at $A=1$. These directions lie within a few degrees of the line $\xi_{\nu_\mu}=\xi_{\nu_\tau}$, where two eigenvalues of the asymmetry matrix meet, so the momentum-resolved code is near the edge of its validity. At $21$ directions near the shifted valleys at $A=1$ and $2$ we ran its version with approximate collisions. It shows the same erratic dependence, and the two codes differ in sign at $6$ of them.
Correcting the momentum-averaged code (Sec.~\ref{sec:codes}) changes $\xi_{\nu_e}^{\rm BBN}$ at these directions by up to $0.04$, against less than $1\%$ at generic directions with $A=0.25$.
At $A=1$ the only direction that both codes allow without the code difference as a theory error is $\phi=337^\circ$. At $A=2$ no direction we computed is allowed by both codes.
We therefore regard the valleys at large asymmetry as not resolved by our codes. Reference~\cite{Domcke:2025lzg} found unexplained disagreements between momentum-averaged and momentum-resolved results in nearby directions.

\textit{Tauphobic line.} For tauphobic asymmetries the crossing exchanges asymmetry between $\nu_e$ and $\nu_\mu$. With $\Hlep$ the sign of $\xi_{\nu_e}^{\rm BBN}$ is reversed at every amplitude we computed, $A=0.05$ to $1$, and $|\xi_{\nu_e}^{\rm BBN}|$ is two to five times larger than without it.
The $95\%$ confidence bound tightens from $|\xi_{\nu_\mu}|<0.23$ to $0.041$ for $\xi_{\nu_\mu}>0$ and stays near $0.065$ for $\xi_{\nu_\mu}<0$, the sign that a first-order QCD transition needs~\cite{Gao:2021nwz}. The criteria of Ref.~\cite{Domcke:2025lzg} give $0.49$ and $0.47$ without $\Hlep$ and $0.17$ and $0.16$ with it.

\section{Sterile-neutrino dark matter}\label{sec:xray}

\textit{X-ray limits.} We use the NuSTAR limits of Ref.~\cite{Krivonos:2024yvm} at $95\%$ confidence for eleven halo profiles, read from the vector graphics of Fig.~3 there. We also use the earlier NuSTAR limits~\cite{Neronov:2016wdd,Perez:2016tcq,Ng:2019gch,Roach:2019ctw,Roach:2022lgo} as compiled in Fig.~10 of Ref.~\cite{Fischer:2022pse}. This compilation ends at $m_s=40\,$keV, so we also read the limits of Refs.~\cite{Perez:2016tcq,Ng:2019gch} over their full ranges, up to $m_s=200\,$keV, from the vector graphics of Fig.~7 of Ref.~\cite{Perez:2016tcq}, for three halo profiles, and of Fig.~4 of Ref.~\cite{Ng:2019gch}. Both give decay rates. Converted to mixing angles, the limit of Ref.~\cite{Ng:2019gch} agrees with Fig.~5 there to within $5\%$ above $m_s=15\,$keV. We take the strongest NuSTAR limit at each mass.
INTEGRAL/SPI gives $3\sigma$ limits, including systematic errors, from $m_s=40.52\,$keV upward for eight halo profiles~\cite{Fischer:2022pse}.
All these limits assume the decay rate of Majorana sterile neutrinos.
Our combined limit agrees with the X-ray region drawn in Ref.~\cite{Akita:2025txo} to within $20\%$ between $m_s=15$ and $30\,$keV, and differs by up to a factor of two in narrow features up to $m_s=39\,$keV.

\textit{Mass limits.} We computed the angle at $m_s=5$, $10$, $20$, $22$, $25$, $30$, $40$, $50$, $60$ and $70\,$keV with $\Hlep$, and at the same masses except $22$ and $30\,$keV without it. Between them we interpolate linearly in $\log m_s$ and $\log\sin^22\theta$.
The mass limit is the lightest mass above $m_s=5\,$keV at which the angle reaches an X-ray limit.
Table~\ref{tab:mlimit} gives it with the observed limits over the halo profiles, and with the median expected limit of Ref.~\cite{Krivonos:2024yvm} for its reference profile, which the Letter quotes. The earlier NuSTAR analyses give no expected limits, so the expected column omits them. The expected limit of Ref.~\cite{Krivonos:2024yvm} ends at $m_s=36.8\,$keV.
With $\Hlep$ the observed limits give $m_s=16.0$ to $16.2\,$keV, within $0.1\,$keV of the flavor-universal values.
Without $\Hlep$ the angle first touches them at $m_s=24.7\,$keV for the strongest profiles, where the observed limit of Ref.~\cite{Krivonos:2024yvm} lies about five times below its median expectation. For the other profiles it first touches them at $m_s=36.1\,$keV, on a single straight segment of the drawn compilation between $m_s=33.9$ and $37.4\,$keV.
With the observed limit of Ref.~\cite{Krivonos:2024yvm} alone, SPI sets the mass limit, at $m_s=51.6\,$keV for the reference profile.

\textit{Variations.} Scaling the computed angles by $0.7$ to $1.3$ moves the mass limit with $\Hlep$ to between $m_s=15.9$ and $18.6\,$keV, the upper end for the weakest profile. For Dirac sterile neutrinos the decay rate is half as large, which is equivalent to halving the angle, and the mass limit becomes $m_s=21.6\,$keV for the reference profile.
SPI sets the mass limits near $m_s=50\,$keV and NuSTAR those below $m_s=40\,$keV, so the two cases are compared at different confidence levels. The SPI limit is the mean line flux plus three standard errors~\cite{Fischer:2022pse}. Scaling it by $1.645/3$ gives the $95\%$ limit only where the mean flux vanishes. If the mean flux is not negative, the $95\%$ limit lies between the scaled and the published limits. The scaled limits lower the masses set by SPI by $2$ to $3\,$keV, so in that case a $95\%$ treatment of SPI would lower them by at most this amount.
The last columns of Table~\ref{tab:mlimit} give the largest allowed mass within the computed range. With $\Hlep$ and our likelihood it is $m_s=19.9$ to $21.0\,$keV, as for flavor-universal asymmetries. Above the observed mass limit the angle stays allowed only in windows narrower than $1\,$keV, and the limits of Refs.~\cite{Perez:2016tcq,Ng:2019gch} exclude it near $m_s=40\,$keV, where the other NuSTAR limits end and SPI begins. With the criteria a window remains between $m_s=40.8$ and $41.9\,$keV. With the expected limit alone the angle stays allowed between $m_s=36.8\,$keV, where that limit ends, and the start of SPI at $m_s=40.52\,$keV. The observed limits of Refs.~\cite{Perez:2016tcq,Ng:2019gch} exclude this range.

\textit{Other directions.} With zero total lepton number the potential that drives the production for mixing with $\nu_\mu$ is proportional to $2L_\mu+L_e+L_\tau=L_\mu$ to leading order~\cite{Akita:2025txo}, so mainly $|L_\mu|$ matters. At the same $|L_\mu|$ our valley runs give angles within $25\%$ of those along $L_\mu-L_\tau$. On our grid of generic directions in the plane, the largest $|L_\mu|$ that BBN allows with $\Hlep$ is $0.004$. In the valley at $A=0.75$ and $\phi=158^\circ$ (Table~\ref{tab:valleys}), the largest amplitude at which the valleys are resolved, it is $0.0086$. We computed the DM angle there at $m_s=5$ to $50\,$keV. The mass limit is $m_s=27\,$keV with the expected NuSTAR limit and $m_s=22.1$ to $23.7\,$keV with the observed limits, against $m_s=52\,$keV along $L_\mu-L_\tau$ without $\Hlep$. The valleys at $A\ge1$ are not resolved (Sec.~\ref{sec:plane}).

\begin{table}[t]
\caption{Sterile-neutrino mass limits in keV for mixing with $\nu_\mu$. The observed column gives the lightest mass above $m_s=5\,$keV at which the angle reaches the observed X-ray limits, over the halo profiles. The expected column uses the median expected NuSTAR limit of Ref.~\cite{Krivonos:2024yvm} for its reference profile and the geometric mean of the SPI limits over the profiles. The last two columns give the largest allowed mass within the computed masses, over the profiles, with SPI at $3\sigma$ and scaled by $1.645/3$. Likelihood and criteria denote the bounds from our likelihood and from the criteria of Ref.~\cite{Domcke:2025lzg}, each for the weaker sign.}
\label{tab:mlimit}
\begin{tabular}{lcccc}
\hline
input & observed & expected & largest allowed & largest, SPI $\times1.645/3$\\
\hline
$L_\mu-L_\tau$, without $\Hlep$, likelihood & $24.7$--$36.1$ & $51.6$ & $48.5$--$52.5$ & $46.3$--$50.0$\\
$L_\mu-L_\tau$, without $\Hlep$, criteria & $50.3$--$53.0$ & $52.5$ & $50.3$--$52.9$ & $47.4$--$51.3$\\
$L_\mu-L_\tau$, with $\Hlep$, likelihood & $16.0$--$16.2$ & $17.8$ & $19.9$--$21.0$ & $19.9$--$21.0$\\
$L_\mu-L_\tau$, with $\Hlep$, criteria & $21.7$--$22.2$ & $23.2$ & $41.2$--$41.9$ & $21.7$--$41.5$\\
flavor-universal, likelihood & $16.0$--$16.3$ & $17.8$ & $19.9$--$21.1$ & $19.9$--$21.1$\\
flavor-universal, criteria & $21.7$--$22.1$ & $23.1$ & $41.1$--$41.8$ & $21.7$--$41.4$\\
\hline
\end{tabular}
\end{table}

\textit{Structure formation.} Reference~\cite{Akita:2025txo} marks, as an illustrative guide and not as a bound, where Lyman-$\alpha$ data may matter. It maps the thermal-relic limit $m_{\rm WDM}>3.1\,$keV onto the sterile-neutrino spectrum by its Eq.~(F3), $m_s<7.56\,{\rm keV}\,(m_{\rm WDM}/3\,{\rm keV})^{4/3}\langle p^2\rangle_0^{1/2}/T_0$. We evaluated it with the final spectra of their code. Along $L_\mu-L_\tau$ at our bound with $\Hlep$, $|L_\mu|=0.0023$, the spectrum has $\langle p\rangle_0/T_0=1.82$ to $1.88$ and the guide lies at $m_s=16.4$ to $16.9\,$keV, from runs at $m_s=16$ and $10\,$keV made for this purpose. At the bound without $\Hlep$, $|L_\mu|=0.032$, the spectrum has $\langle p\rangle_0/T_0=2.01$ at $m_s=16\,$keV and the guide lies at $m_s=18.0\,$keV. At this mass the two runs differ only in $|L_\mu|$, so the smaller asymmetry allowed with $\Hlep$ gives the colder spectrum and the lower guide. A spectrum that follows the active neutrinos gives $m_s=20.3\,$keV, the usual mapping for nonresonant production.

\textit{Depletion by production.} We apply the bounds to the asymmetry before sterile-neutrino production. Production with $\nu_\mu$ mixing depletes the universal asymmetry by $1.5$ to $16\%$ between $m_s=50$ and $5\,$keV, and the asymmetries along $L_\mu-L_\tau$ by at most $5\%$. Rerunning the momentum-averaged code with $\Hlep$ on the depleted inputs at $m_s=5\,$keV lowers $\xi_{\nu_e}^{\rm BBN}$ by $6.9\%$ for the universal asymmetry and raises $|\xi_{\nu_e}^{\rm BBN}|$ by $2.5\%$ along $L_\mu-L_\tau$. Both changes strengthen the conclusion of the Letter.

\end{document}